\documentclass[%
 reprint,
amsmath,amssymb,
aps,
prc,
showpacs,
longbibliography]{revtex4-1}

\usepackage{graphicx}
\usepackage{epsfig} 
\usepackage{dcolumn}
\usepackage{bm}

\begin{document}

\title{Systematic investigation of the Hoyle-analog states in light nuclei.
}
\author{V. S. Vasilevsky}
\email{vsvasilevsky@gmail.com}
\affiliation{Bogolyubov Institute for Theoretical Physics,\\
 Kiev 03143, Ukraine
}
\author{K. Kat\=o}
\email{kato-iku@gd6.so-net.ne.jp}
\affiliation{
 Nuclear Reaction Data Centre, Faculty of Science, Hokkaido University,\\
Sapporo 060-0810, Japan
}
\author{N. Takibayev}
\email{takibayev@gmail.com}

\affiliation{
 Al-Farabi Kazakh National University, \\
Almaty 050040, Kazakhstan
}



\date{\today }

\begin{abstract}
We investigate resonance states in three-cluster continuum of some light
nuclei $^{9}$Be, $^{9}$B, $^{10}$B, $^{11}$B and $^{11}$C. These nuclei are
considered to have a three-cluster configuration consisting of two alpha-particles
and  neutron, proton, deuteron, triton and nucleus $^{3}$He. In this study, we
make use two different microscopic three-cluster  models. The first model
employs the Hyperspherical Harmonics basis to numerate channels and describe
three-cluster continuum. The second model is the well-known complex
scaling method. The nucleon-nucleon interaction is modeled by the semi-realistic
Minnesota and Hasegawa-Nagata potentials. Our main aim is to find the
Hoyle-analog states in these nuclei or, in other words, whether it is possible
to synthesize these nuclei in a triple collision of clusters. We formulate the
criteria for selecting such states and apply  them to resonance states,
emerged from our calculations. We found that there are resonance states
obeying the formulated criteria which make possible syntheses of these nuclei
in a stellar environment.

\end{abstract}

\maketitle

\section{Introduction}

We are going to search and analyze properties of the
Hoyle-like states in light nuclei. It is necessary to recall that the Hoyle
state is a very narrow resonance state in $^{12}$C, which was predicted by
Fred Hoyle in 1954 \cite{1954ApJS....1..121H}. Three years later this state
was experimentally observed by studying beta decays of $^{12}$B in
Ref. \cite{1957PhRv..107..508C}. It is interesting to point out that F. Hoyle
predicted the energy of the $0^{+}$ resonance state at $E$ =0.33 MeV above the
three alpha-particles threshold, and Cook \textit{et al} in Ref.
\cite{1957PhRv..107..508C} determined the position of the resonance state at
$E$= 0.372$\pm$0.002MeV. One has to compare to the modern value of the energy
which is $E$=0.3796$\pm$0.0002 MeV \cite{KELLEY201771}. This resonance state
created by a triple collision of three alpha-particles is the key element in
syntheses of atomic nuclei starting from $^{12}$C. The Hoyle state is a way
for the nucleosynthesis of carbon in helium-burning red giant stars, which are
rich of alpha-particles. Actually, F. Hoyle was the first who proclaimed that
nuclear synthesis can take place in a triple collision of light nuclei, namely
alpha-particles. Such processes are very difficult to be organized in
laboratory, but the Nature has time and facilities to carry out such processes
in the inside of stars. One can find more interesting historical facts and scientific
results about the Hoyle state in the review \cite{2014PrPNP..78....1F}.

Two important quotations from F. Hoyle paper \cite{1954ApJS....1..121H}:

\begin{enumerate}
\item "It was pointed out some years ago by H. Bethe
\cite{1939PhRv...55..434B} that effective element-building inside starts must
proceed, in the absence of hydrogen, by triple collisions as a starting point:%
\begin{equation}
3\alpha\rightarrow^{12}\text{C}+\gamma." \label{eq:H1}%
\end{equation}

\item "It is convenient to replace reaction (\ref{eq:H1}) by%
\begin{equation}
\alpha+\alpha\rightarrow^{8}Be,\quad^{8}Be+\alpha\rightarrow^{12}%
C+\gamma . \label{eq:H2}%
\end{equation}
This is a permissible step, since the lifetime of the unstable $^{8}$Be is
appreciably longer than the time required for nuclear collision of two
$\alpha$ particles; that is, longer than the $\alpha$ particle radius divided
by the relative velocity".
\end{enumerate}

These two equations (\ref{eq:H1}) and (\ref{eq:H2}) represent two different
ways of excitation of the Hoyle resonance state and two different ways of
syntheses of $^{12}$C. However, in both scenarios the very narrow $0^{+}$
resonance state is the key factor in creation of the carbon 12.

There are a very large number of publications devoted to the $0^{+}$  and
other resonance states in $^{12}$C. Different methods have been used to
determine parameters of the Hoyle state and to shed some light on the nature
of this states and other resonances states, residing in the three-cluster
continuum in $^{12}$C. However, only few publications
(\cite{2007PhRvC..75b4302K, 2010PhRvC..82f4315Y,
2011JPhCS.321a2009K, 2012PhRvC..85e4320S,
2012PThPS.196..388Y, 2015PhRvC..91a4316K,
2017arXiv171104439Z, 2018arXiv180100562C}) have been aimed at
finding the Hoyle-analog states in light nuclei. They are mainly concentrated
on closest neighbors of the $^{12}$C nucleus, namely, $^{11}$B, $^{11}$C and
$^{13}$C. In Refs. \cite{2010PhRvC..82f4315Y} and \cite{2012PThPS.196..388Y}\ the
structure of 1/2$^{+}$ and 3/2$^{-}$ states in $^{11}$B has been investigated
within the three-cluster orthogonality condition model (OCM) combined with the Gauss
expansion method. In these papers, parameters of resonance states were
obtained by using the complex scaling technique. By analyzing properties of
wave functions, the authors of the Refs.  \cite{2010PhRvC..82f4315Y,
2012PThPS.196..388Y} came to the conclusion that the 1/2$^{+}$ resonance
state the parameters $E$ = 0.75 MeV and $\Gamma$= 190 keV can be considered as
the Hoyle-analog states. The antisymmetrized molecular dynamics (AMD) has been used
to study the excited states of the negative parity in $^{11}$B and $^{11}$C in
Refs. \cite{2007PhRvC..75b4302K, 2011JPhCS.321a2009K}. It was concluded
that the third excited states in $^{11}$B and $^{11}$C have a dilute cluster
structure $\alpha+\alpha+t$ and $\alpha+\alpha+^{3}$He, respectively, and can
be treated as the Hoyle analog states.

In the present paper we consider these nuclei and also $^{9}$Be, $^{9}$B and
$^{10}$B. We also consider a large number of states with different values of
the total momentum $J$ and both of negative and positive parities. Before
starting in searching for the Hoyle-analog states, one needs to
formulate clear criteria for selecting such states. By analyzing properties of
the Hoyle state, one may suggest the following criteria for the Hoyle analog 
states in three-cluster systems:

\begin{enumerate}
\item Very narrow resonance state,

\item Resonance state which lies close to three-cluster threshold,

\item Resonance state which has the total orbital momentum $L=0$.
\end{enumerate}

We consider the first criterion as the most important as in the case of very
narrow (long-lived) resonance states, and a compound system has more chances
to be reconstructed and transformed in to a bound state. However, we will
analyze all resonance states.

Our main aim is to find the Hoyle-analogue states in light nuclei $^{9}$Be and
$^{9}$B, $^{10}$B, $^{11}$B and $^{11}$C. In other words, we are going to
study whether light nuclei can be created in triple collision of clusters. The
necessary condition for such a process is the existence of a very narrow
resonance state in three-cluster continuum. Actually we consider a chain of
reactions%
\[
A_{1}+A_{2}+A_{3}=A^{\ast}\Rightarrow A+\gamma
\]
which consists of two steps. In the first step, an excited state (very narrow
resonance state) of a compound nucleus is created in a triple collision of
clusters consisting of $A_{1}$, $A_{2}$ and $A_{3}$ nucleons. In the second step, the
compound nucleus by emitting a photon transits from the resonance state to the
bound state. The narrower is a resonance state in the first step, the more is
the probability to transit from the resonance to the bound state. For each
nuclei we determine energy and width of resonance states. We select a
resonance state with a very small width. We also analyze the wave function of
selected resonance states. These investigations will be performed within a
microscopic three-cluster model which involves the hyperspherical harmonics to
distinguish channels of the three-cluster system. For this model, which was
formulated in Ref. \cite{2001PhRvC..63c4606V}, we use the abbreviation AMHHB
which means the algebraic model of scattering making use of the hyperspherical
harmonics basis. In Ref. \cite{2012PhRvC..85c4318V} this model has been applied
to study bound and resonance states in $^{12}$C. It fairly good
reproduced the energy and width of the Hoyle state in $^{12}$C. And it was
demonstrated that this model is in good agreement with other alternative
models, for instance, the complex scaling method. Note that the most
effective methods among others, which are used to study resonance states in
three-cluster and many-channel systems, are the Complex Scaling Method and
Hyperspherical Harmonics Method

We present results obtained with both methods. The AMHHB method, which employs
hyperspherical harmonics to numerate channels of three-cluster continuum,
allows us to determine energy and width of a resonance state, reveals the
dominant decay channels, and sheds more light on the nature of the resonance
state by analyzing its wave functions. This model correctly treats the Pauli
principle and makes uses of the semi-realistic nucleon-nucleon potential.
The complex scaling method (CSM), which also uses this type of the
nucleon-nucleon interaction, is more advance and model independent method to
determine the poles of the  $S$-matrix in two- and three-cluster systems.
 Note  that both methods give very close results for narrow resonance states
and different resonance parameters for wide resonance states.

The preliminary analysis of three-cluster resonance states in $^{9}$B and
$^{9}$B has been carried out in Ref. \cite{PhysRevC.96.034322}, and  
resonance states have been investigated in the
mirror nuclei $^{11}$B and $^{11}$C in Ref.
\cite{2013UkrJPh.58.544V}. In Ref. \cite{2014UkrJPh..59.1065N} the
AMHHB model was applied to study the spectrum of bound states in $^{10}$B. To
make a systematic analysis of resonance states and to discover the Hoyle
analog state in $^{9}$Be, $^{9}$B, $^{10}$B, $^{11}$B and $^{11}$C we have to
make additional calculations and thorough investigations of \ peculiarities of
resonance wave functions.

The present paper is organized in the following way. In Sec. \ref{Sec:Method}
we shortly explain the main idea of the microscopic method which involves the
hyperspherical harmonics for description of bound and scattering states of a
three-cluster system. Results of numerical calculations and discussions of the
results obtained are presented in Sec. \ref{Sec:Results}. We start with the
reexamination of properties of the Hoyle state. We also consider other
resonance states in $^{12}$C to display similarities and differences between
them. This is done within the AMHHB and CSM in order to formulate more clear
criteria for selecting the Hoyle-analog states. And then we proceed with
analysis of resonance states in three-cluster continuum of nuclei $^{9}$Be,
$^{9}$B, $^{10}$B, $^{11}$B and $^{11}$C. By applying the formulated
criteria, we select the Hoyle-analog states and describe their properties.
Sect. \ref{Sec:Conclusion} present a summary of our investigations.

\section{Method \label{Sec:Method}}

\subsection{Three-cluster wave function}

To study three-cluster systems we exploit a microscopic model which
incorporates the Resonating Group Method, the $J$-matrix Method or the
algebraic version of the Resonating Group Method (RGM) and the Hyperspherical
Harmonics Method. Details of the model and its application to study of bound
and continuous spectrum states of light nuclei can be found in Refs.
\cite{2001PhRvC..63c4606V, 2001PhRvC..63c4607V,
2007JPhG...34.1955B, 2010PPN....41..716N,
2012PhRvC..85c4318V, 2013UkrJPh.58.544V} and
\cite{2014PAN..77.555N}.

The standard ansatz of the RGM for representing the wave function of a
three--$s$-cluster system is used
\begin{eqnarray}
\Psi_{E,J} &= & \sum_{S,L}\widehat{\mathcal{A}}\left\{  \left[  \Phi_{1}\left(
A_{1}\right)  \Phi_{2}\left(  A_{2}\right)  \Phi_{3}\left(  A_{3}\right)
\right]  _{S} \right. \label{eq:001} \\%
&\times & \left. \psi_{E,LJ}\left(  \mathbf{x},\mathbf{y}\right)  \right\}  _{J}, \nonumber
\end{eqnarray}
where the wave function $\psi_{E,LJ}\left(  \mathbf{x},\mathbf{y}\right)  $
describes relative motion of clusters \ and the antisymmetric functions
$\Phi_{\nu}\left(  A_{\nu}\right)  $ ($\nu$=1, 2, 3) describes internal motion
of nucleons inside the cluster with index $\nu$. Two vectors $\mathbf{x}$ and
$\mathbf{y}$ denote one of the possible sets of the Jacobi vectors. Within
this paper, the vector $\mathbf{x}$ determines distance between two selected
clusters, while the vector $\mathbf{y}$ represents displacement of the third
cluster with respect to the center of mass of two selected clusters. The
antisymmetrization operator $\widehat{\mathcal{A}}$ provides \ full
antisymmetrization of the wave function of a compound system. By assuming
$\widehat{\mathcal{A}}=1$ and the orthogonality condition to the
Paul-forbidden states, one transits to the OCM.

It is very convenient to use the $LS$ coupling scheme for three interacting
$s$-clusters. In this scheme, the total spin $S$ is a vector sum of individual
spins of clusters, and the total orbital momentum $L$ is also a vector sum of
the partial orbital momenta $\widehat{\mathbf{l}}_{x}$ and $\widehat
{\mathbf{l}}_{y}$, associated with the Jacobi vectors $\mathbf{x}$ and
$\mathbf{y}$, respectively. The total angular momentum $J$ is a vector sum of
the total orbital momentum $L$ and the total spin $S$.

To simplify of obtaining wave functions of discrete and continuous spectrum
states and scattering parameters, we transit from the Jacobi vectors
$\mathbf{x}$ and $\mathbf{y}$ to the hyperspherical coordinates which consist
of hyperradius $\rho$ and five hyperspherical angles which we denote as
$\Omega_{5}$. The hyperradius $\rho$ is defined in unambiguous way
\begin{equation}
\rho=\sqrt{\mathbf{x}^{2}+\mathbf{y}^{2}}, \label{eq:002A}%
\end{equation}
while there are several different ways for definition of the hyperspherical
angles (see for instance, \cite{1965AnPhy..35...18Z, Avery89,
Dzhibuti84E}). We make use the most popular set of hyperspherical angles
which was suggested by Zernike and Brinkman in 1935 \cite{kn:ZB}.
This set consists of the hyperspherical angle $\theta$ which determines
relative lengths of the Jacobi vectors
\begin{equation}
x=\rho\cos\theta,\quad y=\rho\sin\theta, \label{eq:002B}%
\end{equation}
two angles $\theta_{x}$and $\phi_{x}$, determining orientation of vector
$\mathbf{x}$\textbf{, }and two other angles $\theta_{y}$ and $\phi_{y}$,
determining orientation of vector $\mathbf{y}$ in the space. Note, that the
angles $\left\{  \theta_{x},\phi_{x}\right\}  $ describe rotation of a
two-cluster subsystem and the angles $\left\{  \theta_{y},\phi_{y}\right\}  $
describe rotation of the third cluster around center of mass of the
two-cluster subsystem. Five hyperspherical angles are able to describe any
shape and any orientation (i.e. rotation) of a triangle connecting centers of
mass of three clusters, and hyperradius determines any size of that triangle.

Having introduced the hyperspherical coordinate, we can represent the
three-cluster wave function (\ref{eq:001}) in the following form
\begin{align}
\Psi_{E,J}  &  =\sum_{c}\widehat{\mathcal{A}}\left\{  \left[  \Phi_{1}\left(
A_{1}\right)  \Phi_{2}\left(  A_{2}\right)  \Phi_{3}\left(  A_{3}\right)
\right]  _{S}\right. \label{eq:003}\\
&  \times\left.  \psi_{E,c}\left(  \rho\right)  \mathcal{Y}_{c}\left(
\Omega_{5}\right)  \right\}  _{J},\nonumber
\end{align}
where $c$ is a multiple index $c=\left\{  K;\lambda,l;L,S\right\}  $
classifying channels of the three-cluster system and involving the
hypermomentum $K$, partial orbital momenta $\lambda$\ and $l$ associated with
the Jacobi vectors $\mathbf{x}$ and $\mathbf{y}$, respectively, and the total
orbital momentum $L$. The hyperspherical harmonics $\mathcal{Y}_{c}\left(
\Omega_{5}\right)  $ form a complete set of functions on five-dimension sphere
and thus account for all kinds of motion of a three-cluster system. Components
of the many-channel hyperradial wave function $\left\{  \psi_{E,c}\left(
\rho\right)  \right\}  $ have to be determined by solving the Schr\"{o}dinger
equation with the selected nucleon-nucleon potential.

\subsection{Three-cluster equation}

For three structureless particles one obtains the infinite set of differential
equations%
\begin{equation}
\sum_{\widetilde{c}}\left[  \delta_{c,\widetilde{c}}\widehat{T}_{K}%
+V_{c,\widetilde{c}}\left(  \rho\right)  \right]  \psi_{E,\widetilde{c}%
}\left(  \rho\right)  =E\psi_{E,c}\left(  \rho\right)  , \label{eq:008}%
\end{equation}
where%
\begin{equation}
\widehat{T}_{K}=-\frac{\hbar^{2}}{2m}\left[  \frac{\partial^{2}}{\partial
\rho^{2}}+\frac{5}{\rho}\frac{\partial}{\partial\rho}-\frac{K\left(
K+4\right)  }{\rho^{2}}\right]  . \label{eq:008A}%
\end{equation}
Matrix $\left\Vert V_{c,\widetilde{c}}\left(  \rho\right)  \right\Vert $ of
the effective potential energy is determined as matrix elements of interaction
$\widehat{V}$ between the hyperspherical harmonics%
\begin{equation}
V_{c,\widetilde{c}}\left(  \rho\right)  =\left\langle \mathcal{Y}%
_{c}\left\vert \widehat{V}\right\vert \mathcal{Y}_{\widetilde{c}}\right\rangle
, \label{eq:009}%
\end{equation}
where integration is performed over all hyperspherical angles $\Omega_{5}$. If
particles have electric charges, than we have the following contribution
\begin{equation}
V_{c,\widetilde{c}}^{\left(  C\right)  }\left(  \rho\right)  =\frac
{Z_{c,\widetilde{c}}e^{2}}{\rho} \label{eq:009A}%
\end{equation}
from the Coulomb interaction to the potential energy $V_{c,\widetilde{c}%
}\left(  \rho\right)  $ (\ref{eq:009}). The quantity $Z_{c,\widetilde{c}}$ can
be called the effective charge. Assuming that at a large values of hyperradius  the
effective potential $V_{c,\widetilde{c}}\left(  \rho\right)  $ originated from
a short range particle-particle interaction is negligibly small, and omitting
non-diagonal elements of the effective charge (that is putting $Z_{c,\widetilde
{c}}=0$ for $c\neq\widetilde{c}$), we obtain an asymptotic part of the channel
Hamiltonian%
\begin{equation}
\widehat{H}_{c}^{\left(  A\right)  }=\left\{  -\frac{\hbar^{2}}{2m}\left[
\frac{\partial^{2}}{\partial\rho^{2}}+\frac{5}{\rho}\frac{\partial}%
{\partial\rho}-\frac{K\left(  K+4\right)  }{\rho^{2}}\right]  +\frac
{Z_{c,c}e^{2}}{\rho}\right\}  . \label{eq:011}%
\end{equation}
Eigenfunctions of this Hamiltonian describing incoming and outgoing
hyperradial waves can be easily found and expressed through the Whittaker
functions (see chapter 13.1 in Ref. \cite{kn:abra})
\begin{equation}
\psi_{c}^{\left(  \pm\right)  }\left(  \rho,\eta_{c}\right)  =\sqrt{\frac{\pi
}{2}}\frac{1}{\rho^{5/2}}W_{\mp i\eta_{c},K+2}\left(  \mp2ik\rho\right)  ,
\label{eq:012}%
\end{equation}
where%
\[
k=\sqrt{\frac{2mE}{\hbar^{2}}}%
\]
and $\eta_{c}$ is the Sommerfeld parameter for the three-cluster system%
\[
\eta_{c}=\frac{m}{\hbar^{2}}\frac{Z_{c,c}e^{2}}{k}.
\]
Thus, the boundary conditions or the asymptotic form of many-channel wave
functions can be expressed in the form%
\[
\psi_{E,c}\left(  \rho\right)  =\delta_{c_{0},c}\psi_{c}^{\left(  -\right)
}\left(  \rho,\eta_{c}\right)  -S_{c_{0},c}\psi_{c}^{\left(  +\right)
}\left(  \rho,\eta_{c}\right)  ,
\]
where $c_{0}$ stands for an incoming channel, $S_{c_{0},c}$\ is an element of
the scattering $S$-matrix.

For three-cluster systems, when the internal structure of clusters and the
Pauli principle are taking into account, we obtain the set of
integro-differential equations:%

\begin{eqnarray}
&  \sum_{\widetilde{c}}\left[  \delta_{c,\widetilde{c}}\widehat{T}_{K}%
\psi_{E,\widetilde{c}}\left(  \rho\right)  +\int d\widetilde{\rho}%
\widetilde{\rho}^{5}V_{c,\widetilde{c}}\left(  \rho,\widetilde{\rho}\right)
\psi_{E,\widetilde{c}}\left(  \widetilde{\rho}\right)  \right]  \label{eq:020}%
\\
&  =E\sum_{\widetilde{c}}\int d\widetilde{\rho}\widetilde{\rho}^{5}%
N_{c,\widetilde{c}}\left(  \rho,\widetilde{\rho}\right)  \psi_{E,\widetilde
{c}}\left(  \widetilde{\rho}\right)  .\nonumber
\end{eqnarray}

This system of equations can be obtained from the many-particle Schr\"{o}dinger
equations with the help of the projection operator%
\begin{equation}
\widehat{P}_{c}\left(  \rho\right)  =\widehat{\mathcal{A}}\left\{  \left[
\Phi_{1}\left(  A_{1}\right)  \Phi_{2}\left(  A_{2}\right)  \Phi_{3}\left(
A_{3}\right)  \right]  _{S}\delta\left(  \rho-\overline{\rho}\right)
\mathcal{Y}_{c}\left(  \Omega_{5}\right) . \right\} \label{eq:021}%
\end{equation}
Applying this operator to the unit operator, we obtain the norm kernel
$N_{c,\widetilde{c}}\left(  \rho,\widetilde{\rho}\right)  $
\begin{equation}
N_{c,\widetilde{c}}\left(  \rho,\widetilde{\rho}\right)  =\left\langle
\widehat{P}_{c}\left(  \rho\right)  |\widehat{P}_{\widetilde{c}}\left(
\widetilde{\rho}\right)  \right\rangle . \label{eq:022}%
\end{equation}
In this expression integration is performed over all spacial coordinates (the
Jacobi vectors) and over all spin and isospin coordinates as well. The matrix
of the potential energy is related to matrix elements of the microscopic
Hamiltonian $\widehat{H}$ by the relation
\begin{equation}
V_{c,\widetilde{c}}\left(  \rho,\widetilde{\rho}\right)  =\left\langle
\widehat{P}_{c}\left(  \rho\right)  \left\vert \widehat{H}\right\vert
\widehat{P}_{\widetilde{c}}\left(  \widetilde{\rho}\right)  \right\rangle
-\delta_{c,\widetilde{c}}\widehat{T}_{K}\delta\left(  \rho-\widetilde{\rho
}\right)  . \label{eq:023}%
\end{equation}
The system of Eq. (\ref{eq:020}) can be directly solved by reducing to the
reasonable finite number of involved three-cluster channels $N_{c}$ and with
the boundary conditions determined above. Solutions of the systems yields us
the definite set of matrix elements of the $S$ matrix. They describe all
kinds of elastic and inelastic processes in a three-cluster system.

Note that the operator (\ref{eq:021}) is a straightforward generation of the
projection operator which has been used for two-cluster systems (see Ref.
\cite{1978PhR....47..167T}). In three-cluster systems, we can easily perform 
this operation though we do not explain the details here.

Within the present model a wave function (\ref{eq:001}) of a three-cluster
system is expanded over an infinite set of cluster oscillator functions
$\left\vert n_{\rho},c\right\rangle $%
\[
\Psi_{E,LJ}=\sum_{n_{\rho},c}C_{n_{\rho},c}^{E,J}\left\vert n_{\rho
},c\right\rangle ,
\]
where%
\begin{eqnarray}
&& \left\vert n_{\rho},c\right\rangle    =\left\vert n_{\rho},K;\lambda
,l;L\right\rangle \label{eq:010}\\
&  =& \widehat{\mathcal{A}}\left\{  \Phi_{1}\left(  A_{1}\right)  \Phi
_{2}\left(  A_{2}\right)  \Phi_{3}\left(  A_{3}\right)  R_{n_{\rho}K}\left(
\rho,b\right)  \mathcal{Y}_{c}\left(  \Omega_{5}\right)  \right\}  ,\nonumber
\end{eqnarray}
 $\mathcal{Y}_{c}\left(  \Omega_{5}\right)  $\ is a hyperspherical
harmonic with the quantum numbers $c=\left\{  K,l_{x},l_{y},L\right\}  $ and
$R_{n_{\rho},K}\left(  \rho,b\right)  $ is an oscillator function%
\begin{eqnarray}
R_{n_{\rho},K}\left(  \rho,b\right)   &  =&\left(  -1\right)  ^{n_{\rho}%
}\mathcal{N}_{n_{\rho},K}\label{eq:010A}\\
& \times &      r^{K}\exp\left\{  -\frac{1}{2}r^{2}\right\}
L_{n_{\rho}}^{K+3}\left(  r^{2}\right)  ,                        \nonumber \\
r &  =&\rho/b,\quad\mathcal{N}_{n_{\rho},K}=b^{-3}\sqrt{\frac{2\Gamma\left(
n_{\rho}+1\right)  }{\Gamma\left(  n_{\rho}+K+3\right)  }},\nonumber
\end{eqnarray}
and $b$ is an oscillator length.

In this case, a set of the integro-differential equations is reduced to a set
of the algebraic (matrix) equations%

\begin{equation}
\sum_{\widetilde{n}_{\rho},\widetilde{c}}\left[  \left\langle n_{\rho
},c\left\vert \widehat{H}\right\vert \widetilde{n}_{\rho},\widetilde
{c}\right\rangle -E\left\langle n_{\rho},c|\widetilde{n}_{\rho},\widetilde
{c}\right\rangle \right]  C_{\widetilde{n}_{\rho},\widetilde{c}}^{E,J}=0,
\label{eq:030}%
\end{equation}
which can be more easily solved by the numerical methods than the set of
equations (\ref{eq:020}). For continuous spectrum states one has to impose
proper boundary conditions for expansion coefficients $\left\{  C_{n_{\rho}%
,c}^{E,J}\right\}  $. These conditions have been discussed in Ref.
\cite{2001PhRvC..63c4606V} where relations between the discrete $\left\{
C_{n_{\rho},c}^{E,J}\right\}  $ and continuous $\left\{  \psi_{E,c}\left(
\rho\right)  \right\}  $\ wave functions were established. By including the
asymptotic form of expansion coefficients $\left\{  C_{n_{\rho},c}%
^{E,J}\right\}  $, which is valid for large values of hyperradial excitations
$n_{\rho}\gg1$, we obtain in a closed form the system of equations determining
both wave functions of a continuous spectrum and the corresponding $S$ matrix.

\subsection{Supplementary quantities}

Having obtained the expansion coefficients for any state of the three-cluster
continuum, we can easily construct its wave function in the coordinate space.
It can be done, the first of all, for the total hyperradial wave function
\begin{equation}
\psi_{E,c}\left(  \rho\right)  =\sum_{n_{\rho}}C_{n_{\rho},c}^{E,J}R_{n_{\rho
},K}\left(  \rho,b\right).  \label{eq:033}%
\end{equation}
It can be also done for the wave function%
\begin{equation}
\psi_{E,LJ}\left(  \mathbf{x},\mathbf{y}\right)  =\sum_{n_{\rho},c}C_{n_{\rho
},c}^{E,J}R_{n_{\rho},K}\left(  \rho,b\right)  \mathcal{Y}_{c}\left(
\Omega_{5}\right).  \label{eq:034}%
\end{equation}

To get more information about the state under  consideration we will study
different quantities which can be obtained  with the wave function in discrete
or coordinate spaces. With wave functions in the discrete oscillator quantum number
representation we can determine a weight $W_{sh}$ of the oscillator function
belonging to the oscillator shell $N_{sh}$ in this wave function:
\begin{equation}
W_{sh}\left(  N_{sh}\right)  =\sum_{n_{\rho},c\in N_{sh}}\left\vert
C_{n_{\rho},c}^{E,J}\right\vert ^{2}. \label{eq:035}%
\end{equation}
where the summation is performed over all hyperspherical harmonics and 
hyperradial excitations obeying the following condition:
\[
N_{os}=2n_{\rho}+K.
\]
Here $N_{os}$ is fixed. Basis wave functions (\ref{eq:010A}) belongs to the oscillator shell with the
number of oscillator quanta $N_{os}=2n_{\rho}+K$. It is convenient to numerate
the oscillator shells by $N_{sh}$ ( = 0, 1, 2, . . . ), which we determine as
\[
N_{os}=2n_{\rho}+K=2N_{sh}+K_{\min},
\]
where $K_{\min}=L$ for normal parity states $\pi=\left(  -1\right)  ^{L}$ and
$K_{\min}=L+1$ for abnormal parity states $\pi=\left(  -1\right)  ^{L+1}$.
Thus we account oscillator shells starting from a "vacuum" shell ($N_{sh}$ =
0) with minimal value of the hypermomentum $K_{\min}$ compatible with a given
total orbital momentum $L$.

The weights $W_{sh}$ we will calculate both for bound and resonance states. For a
bound state, the wave function is normalized by the condition%
\begin{equation}
\left\langle \Psi_{E,J}|\Psi_{E,J}\right\rangle =\sum_{n_{\rho}%
,c}\left\vert C_{n_{\rho},c}^{E,J}\right\vert ^{2}=1, \label{eq:036A}%
\end{equation}
and this quantity $W_{sh}$ determines the probability. For the continuous spectrum state,
when the wave function is normalized by the condition%
\begin{equation}
\left\langle \Psi_{E,J}|\Psi_{\widetilde{E},J}\right\rangle
=\sum_{n_{\rho},c}C_{n_{\rho},c}^{E,J}C_{n_{\rho},c}^{\widetilde{E},J}%
=\delta\left(  k-\widetilde{k}\right)  , \label{eq:036B}%
\end{equation}
this quantity has a different meaning. It determines the relative contribution
of the different oscillator shells and also the shape of  the resonance wave
function in the oscillator representation.

It is worthwhile to notice that oscillator functions have some important features. 
Oscillator functions belonging to an oscillator shell $N_{sh}$ allow one to 
describe a many-particle system in a finite range of hyperradius 
$0<\rho \le b\sqrt{4N_{sh}+K_{\min}+3}$. Outside this region, these oscillator 
functions give a negligible small contribution to  many-particle wave function. 
This statement is, for example, demonstrated in Ref. \cite{PhysRevC.97.064605}. 
Thus, oscillator functions with  a small value of $N_{sh}$ describe very compact 
configurations of a three-cluster system with all clusters being close to each 
other. When $N_{sh}$ is large,  the oscillator functions represent a dispersed 
(dilute) configurations. There are two principal regimes in these configurations. 
The first regime is associated with a two-body type of asymptotic when two clusters 
are at a small distance and the third cluster is moved far away. The second regime 
accounts for the case when all three clusters are well separated. Taking these into 
account, we will deduce from an analysis of shell weights $W_{sh}$  whether a 
wave function of a bound or resonance state describes a compact or dispersed 
three-cluster configuration.

By employing the wave function in the coordinate space we determine the
correlation function
\begin{equation}
D\left(  x,y\right)  =x^{2}y^{2}\int\left\vert \psi_{E,LJ}\left(
\mathbf{x},\mathbf{y}\right)  \right\vert ^{2}d\widehat{\mathbf{x}}%
d\widehat{\mathbf{y}} \label{eq:037}%
\end{equation}
and average distances  $R_{1}$ and $R_{2}$ between clusters
\begin{eqnarray}
R_{1} &  =\sqrt{\frac{A}{\left(  A_{1}+A_{2}\right)  A_{3}}}\sqrt{\int
y^{2}\left\vert \psi_{E,LJ}\left(  \mathbf{x},\mathbf{y}\right)
\right\vert ^{2}d\mathbf{x}d\mathbf{y}},\label{eq:038A}\\
R_{2} &  =\sqrt{\frac{\left(  A_{1}+A_{2}\right)  }{A_{1}A_{2}}}\sqrt{\int
x^{2}\left\vert \psi_{E,LJ}\left(  \mathbf{x},\mathbf{y}\right)
\right\vert ^{2}d\mathbf{x}d\mathbf{y}}.\label{eq:038B}%
\end{eqnarray}
In our notations, $R_{2}$ determines an average distance between alpha-particles,
while $R_{1}$ determines a distance of the third cluster to the center of mass
of two alpha particles. Note that in Eq. (\ref{eq:037}) integration is
performed over unit vectors $\widehat{\mathbf{x}}\ $and $\widehat{\mathbf{y}}%
$, while in Eqs. (\ref{eq:038A}) and (\ref{eq:038B}) integration is carried
out over all Jacobi vectors or all hyperspherical coordinates.

It is obvious, that the correlation function $D\left(  x,y\right)  $ can be
determined both for bound and resonance states. However, the average distances
$R_{1}$ and $R_{2}$ can be calculated for the bound state only, since \ for
resonance states integrals in Eqs. (\ref{eq:038A}) and (\ref{eq:038B})
diverge. In Ref. \cite{PhysRevC.96.034322} we suggested to extent to resonance
states the definition of average distances $R_{1}$ and $R_{2}$. \ For this aim
we restricted the integration within the internal part of the resonance wave functions
which was normalized to unity. Recall that the internal part of a wave
function is represented in the region (0$\leq\rho\leq\rho_{\max}$ in the
coordinate space or 0$\leq n_{\rho}\leq N^{\left(  i\right)  }$ in the
oscillator space) where distances between clusters are relatively small and
effects intercluster interactions are very strong. Such the definition of
$R_{1}$ and $R_{2}$ allows us to study the shape of the triangle, composed by
three interacting clusters, but not its size. By comparing average distances
$R_{1}$ and $R_{2}$ for different resonances of the same or other nucleus, we
obtain more information on the structure of the resonance wave functions.

It is important to note, that the oscillator basis (\ref{eq:010}) can be used
to determine parameters of resonance states within the methodology of the
complex scaling method. It will be demonstrated in other paper. However, it is
more expedient to use the Gaussian basis in the six-dimension space to perform
such a type of calculations, as this basis provides more rapid convergence of
results than the oscillator basis.

\section{Results and discussions. \label{Sec:Results}}

For all nuclei under consideration we employ the Minnesota potential
(\cite{kn:Minn_pot1, 1970NuPhA.158..529R}) \ (MP) or the modified
Hasegawa-Nagata potential \cite{potMHN1, potMHN2} (MHNP). Both the central and
spin-orbital components of these potentials are taken into account.

In such a type of calculations we have only one free parameter to be selected.
This is the oscillator length $b$ which is common for all clusters of a
compound nucleus and  effectively determines the spatial distribution of
nucleons in clusters. In our calculations the oscillator length $b$ is fixed
by minimizing the energy of the three-cluster threshold. For $^{9}$Be, $^{9}$B
and $^{12}$C, the oscillator length $b$ minimizes the energy of an alpha-particle.

The Majorana parameter $m$ of the MHNP and the exchange parameter $u$ of the MP are
very often used as an adjustable parameter. If one adjusts these parameters to
reproduce phase shifts of the  $\alpha-\alpha$ scattering and parameters of
resonance states in $^{8}$Be, one obtains the overbound the $0^{+}$ and
$2^{+}$ states in $^{12}$C, and also the $4^{+}$ bound state which contradicts
to the experimental data. This problem was discussed in Ref.
\cite{2012PhRvC..85c4318V} where the MP was used to calculate spectrum of
bound and resonance states in $^{12}$C. Such a problem  also appears for all
nuclei under considerations. This problem has been discussed in Refs.
\cite{2014PAN..77.555N, PhysRevC.96.034322} where spectra of $^{9}$B
and $^{9}$Be were investigated. To avoid appearance unphysical bound states,
we adjust parameters $m$ and $u$ to reproduce the energy of the ground state
measured from the three-cluster threshold. For mirror nuclei $^{9}$Be and
$^{9}$B, $^{11}$B and $^{11}$C, we adjust these parameters only for one
nucleus of these pairs; for $^{9}$B and $^{11}$B. This is done in order to
study effects of the Coulomb interaction on parameters of bound and resonance
states. In Table \ref{Tab: InputParam} we collected input parameters for each
nucleus. We also demonstrate the energy ($E_{R}$) and width ($\Gamma_{R}$) of
the 0$^{+}$ resonance state in $^{8}$Be, obtained with these input parameters.%

\begin{table}[tbp] \centering
\begin{ruledtabular}
\caption{List of nuclei investigated within  microscopic three-cluster  models
and input parameters; $b$ - oscillator length, $m$ or $u$ - exchange 
parameters of the NN interaction. $E_R$ (MeV) and $\Gamma_R$ (keV) are 
parameters of the $0^{+}$ resonance state in $^8$Be.}%
\begin{tabular}
[c]{lllllll}
Nucleus & 3CC & Potential & $u$ or $m$ & $b$, fm & $E_{R}$ & $\Gamma_{R}%
$, \\\hline
$^{9}Be$ & $\alpha+\alpha+n$ & MHNP & 0.0332 & 1.317 & 0.859 & 958.4\\
$^{9}B$ & $\alpha+\alpha+p$ & MHNP & 0.0332 & 1.317 & 0.859 & 958.4\\
$^{10}B$ & $\alpha+\alpha+d$ & MP & 0.915 & 1.395 & 0.426 & 69.0\\
$^{11}B$ & $\alpha+\alpha+^{3}H$ & MP & 0.920 & 1.322 & 0.317 & 19.8\\
$^{11}C$ & $\alpha+\alpha+^{3}He$ & MP & 0.920 & 1.322 & 0.317 & 19.8\\\hline
$^{12}C$ & $\alpha+\alpha+\alpha$ & MP & 0.940 & 1.285 & 0.022 &
2.33$\cdot10^{-5}$\\
\end{tabular}
\label{Tab: InputParam}%
\end{ruledtabular}
\end{table}%

One can see that by selecting the optimal values of the parameters $u$ and $m$
of nucleon-nucleon interaction, we make very broad the 0$^{+}$ resonance state
in $^{8}$Be.

Having determined the oscillator length $b$ and the parameter of the
nucleon-nucleon forces, we have to select a part of the total Hilbert
space which takes part in construction of  the wave function of three-cluster
continuous states. This part is restricted by the number of the three-cluster
channels $c$ and the number of hyperradial excitations or, in other words, the
maximal number of oscillator shell. In all our calculations we use a standard
set of the hyperspherical harmonics and hyperradial excitations. Positive
parity states are calculated with the hyperspherical harmonics $K_{\min}\leq
K\leq K_{\max}$. where $K_{\max}=14$ for the positive parity states and
$K_{\max}=13$ for the negative parity states. The minimal value of the
hypermomentum $K_{\min}$ equals the total orbital momentum $L$ for normal
parity states $\pi=\left(  -1\right)  ^{L}$ and $K_{\min}=L+1$ for the
non-normal parity states. The total number of channels \ $N_{ch}$ depends on
the total angular momentum $J$, the possible values of the total orbital
moment $L$ and symmetry properties of a three-cluster system. To achieve the
asymptotic region and to provide sufficient precision of our calculations we
take into account the hyperradial excitation up to 70. This value of
hyperradial excitations and the number of the hyperspherical channels cover a
large range of intercluster distances and different shapes of the
three-cluster triangle.

In this paper we will not discuss the dependence of parameters of resonance states 
on  $K_{\max}$ and $N_{ch}$, and the convergence of the obtained results, as they were 
addressed in Refs. \cite{2014PAN..77.555N, 2014UkrJPh..59.1065N, 2013UkrJPh.58.544V, 
PhysRevC.96.034322, 2012PhRvC..85c4318V}.

Within our models, the total spin $S$ of odd nuclei $^{9}$Be, $^{9}$B, $^{11}$B
and $^{11}$C equals 1/2, thus the two following values of the total orbital
momentum are involved in calculations:%
\[
L=J-1/2,\quad J=L+1/2.
\]
The total spin $S$\ of the odd-odd nucleus $^{10}$B equals one, therefore
bound and resonance states of the nucleus are constructed by three values of
the total orbital momentum
\[
L=J-1,\quad J=L,\quad J=L+1.
\]
An interesting feature of description of the $^{10}$B within 
the hyperspherical harmonics is that it includes almost two times more 
hyperspherical channels $c$ than in  nuclei $^{9}$Be, $^{9}$B, $^{11}$B
and $^{11}$C.  
Note that the coupling of states with different values of the total 
orbital momentum $L$ is
totally determined by the spin-orbital interaction of nucleons.

\subsection{$^{12}$C:\ Hoyle state}

In this section we are going to reexamine some results obtained in previous
papers concentrating our much interest to properties of the Hoyle state in $^{12}$C.

In Table \ref{Tab:12CHHBvsCSM} we compare parameters of resonance states
obtained within AMHHB \cite{2012PhRvC..85c4318V} and CSM
\cite{2007NuPhA.792...87K}. There is some consistencies in these two different
methods of obtaining resonance states in the three-cluster continuum. Energy
and total width of the first $0^{+}$ resonance state (the Hoyle state) are
very close in both methods. The same is observed for other narrow $1^{-}$
resonance states in $^{12}$C.%

\begin{table}[tbp] \centering
\begin{ruledtabular}
\caption{Spectrum of low-lying resonance states in $^{12}$C calculated
within the AMHHB and CSM.}%
\begin{tabular}
[c]{ccccc}
& \multicolumn{2}{c}{CSM \cite{2007NuPhA.792...87K}} &
\multicolumn{2}{c}{AMHHB \cite{2012PhRvC..85c4318V}}\\\hline
$J^{\pi}$ & $E$, MeV & $\Gamma$, keV & $E$, MeV & $\Gamma$, keV\\\hline
$0^{+}$ & 0.76 & 2.4 & 0.68 & 2.9\\
& 1.66 & 1480 & 5.16 & 534\\
$2^{+}$ & 2.28 & 1100 & 2.78 & 10\\
& 5.14 & 1900 & 3.17 & 280\\
& 6.82 & 240 & 5.60 & 0.6\\
$1^{-}$ & 3.65 & 0.30 & 3.52 & 0.21\\
$3^{-}$ & 1.51 & 2.0$\times$10$^{-3}$ & 0.67 & 8.34\\
\end{tabular}
\label{Tab:12CHHBvsCSM}%
\end{ruledtabular}
\end{table}%

\begin{figure}[ptbh]
\begin{center}
\includegraphics[width=\columnwidth]%
{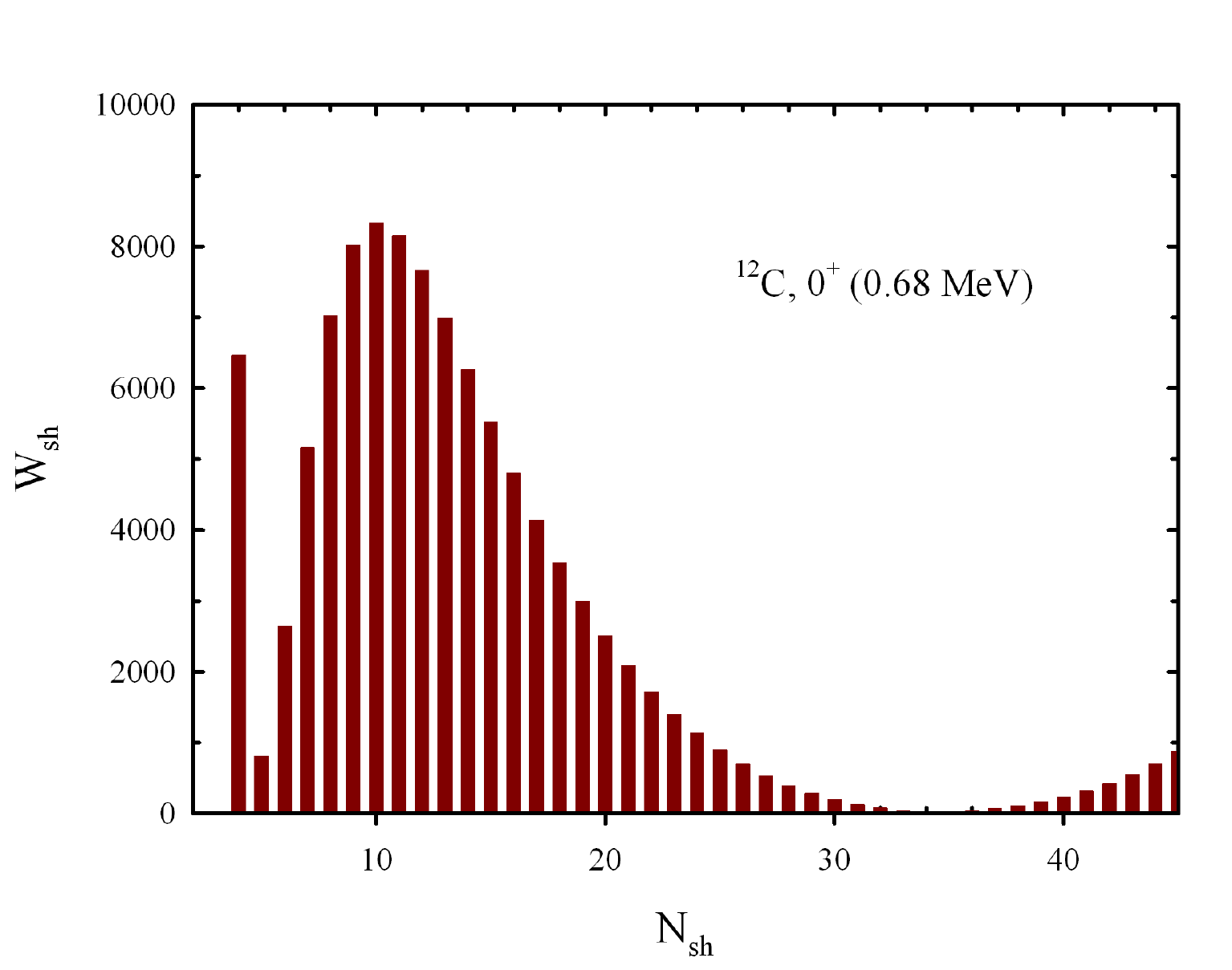}
\caption{Weights of different oscillator shells in the wave function of the
first $0^{+}$ resonance state in $^{12}$C.}%
\label{Fig:ShellWeights0P12C}%
\end{center}
\end{figure}

In Fig. \ref{Fig:ShellWeights0P12C} we display the structure of the wave
function of the Hoyle state. As we see the weights of oscillator shells have
very large amplitudes and main contribution to the wave function in the
internal region comes from the oscillator shells $0\leq N_{sh}\leq30$. In the
asymptotic region this function has an oscillatory behavior with much smaller
amplitude. We consider such a behavior of a resonance wave function as a
'standard' or pattern for the Hoyle analog states.%

\begin{figure}[ptbh]
\begin{center}
\includegraphics[width=\columnwidth]%
{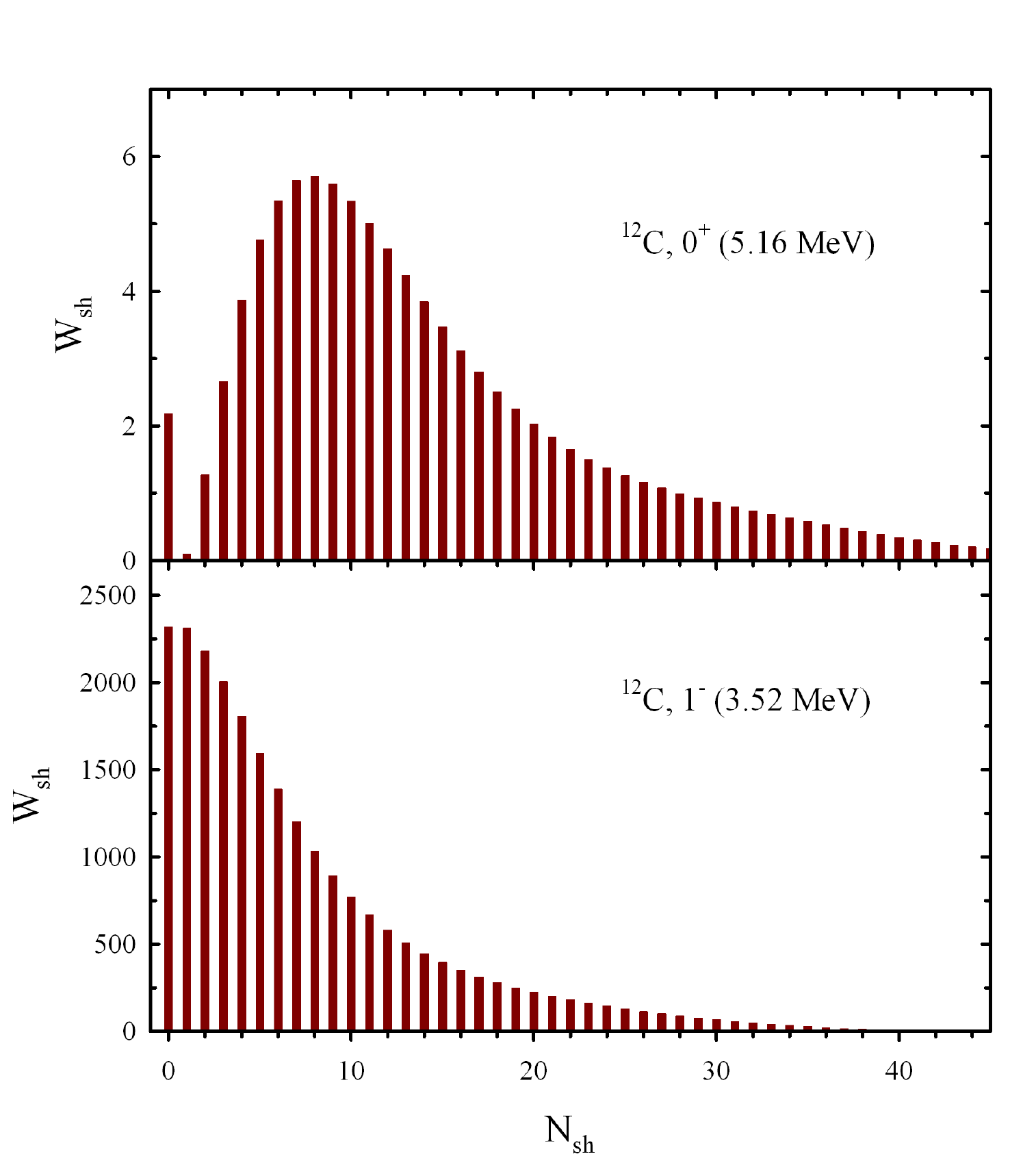}
\caption{Structure of wave functions of the second $0^{+}$ and first $1^{-}$
resonance states in $^{12}$C.}%
\label{Fig:ShellWeight12CRS}%
\end{center}
\end{figure}

It is interesting to compare wave function of the Hoyle state with wave
functions of other resonance states in $^{12}$C. We selected the second
$0^{+}$ resonance state and $1^{-}$ resonance state. As it follows from Table
\ref{Tab:12CHHBvsCSM}, the second $0^{+}$ resonance state is a broad resonance
state ($\Gamma$ = 534 keV) while the $1^{-}$ resonance state is a narrow
resonance state ($\Gamma$ =0.21 keV). Presented wave functions of these two
states (Fig. \ref{Fig:ShellWeight12CRS}) demonstrate that the wave function
of narrow $1^{-}$ state has a behavior which is close to the standard behavior
of the Hoyle state, as it has very large amplitudes of the oscillator shells
$0\leq N_{sh}\leq30$. Contrary to this case, the wave function of the second
resonance $0^{+}$ state has rather small amplitudes of the lowest oscillator
shells. It is naturally to assume that the $1^{-}$ resonance state is the
Hoyle-analog state in $^{12}$C. We will use the standard behavior of the wave
function of the Hoyle state, displayed in Fig. \ref{Fig:ShellWeights0P12C},
as the additional criterion for selecting the Hoyle-analog states.

\begin{figure}[ptbh]
\begin{center}
\includegraphics[width=\columnwidth]%
{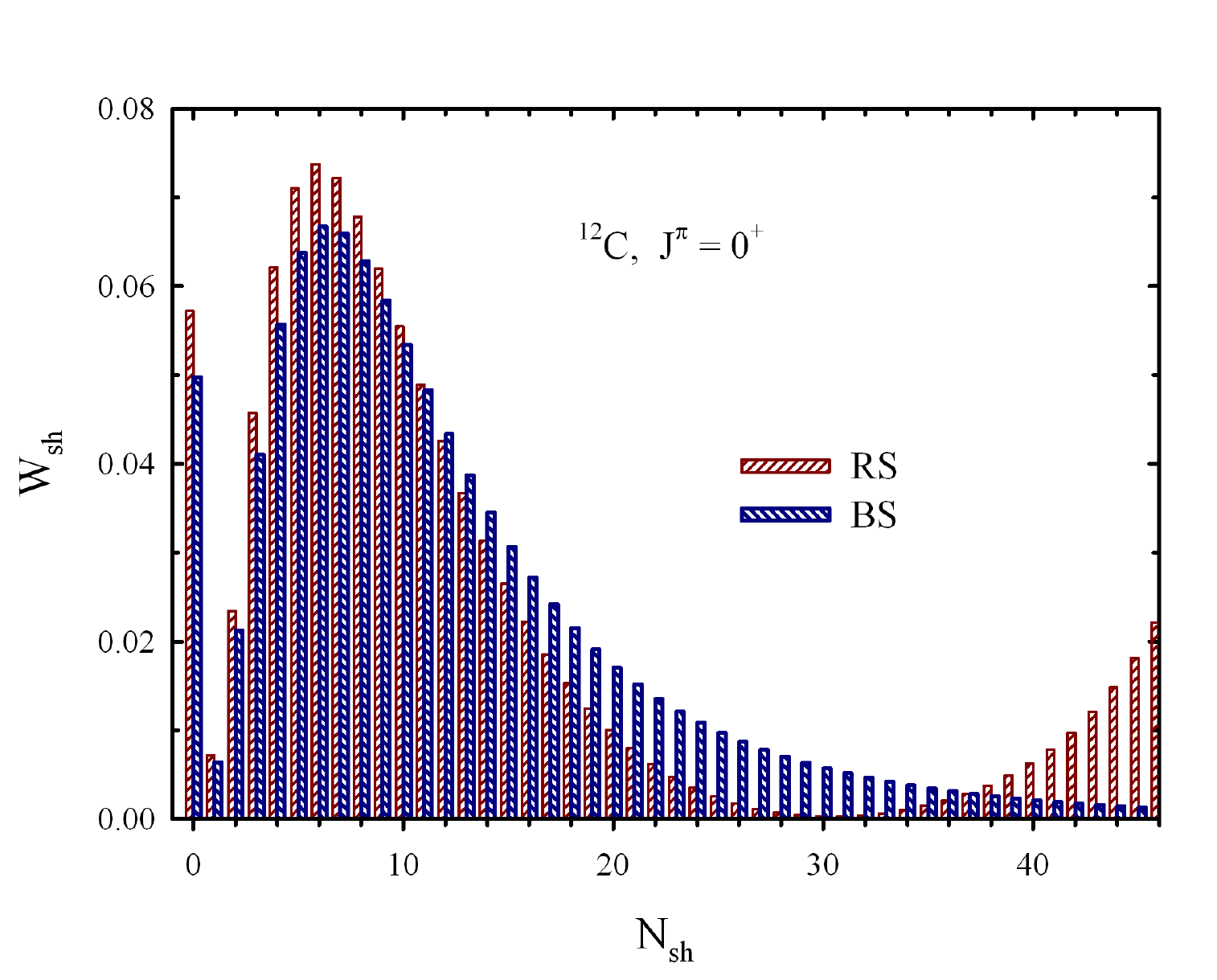}
\caption{Comparing wave functions of the $0^{+}$ resonance state (RS) and the $0^{+}$
pseudo-bound (BS) state in $^{12}$C.}%
\label{Fig:BSvsRS0P12C}%
\end{center}
\end{figure}

In Fig. \ref{Fig:BSvsRS0P12C} we compare the resonance wave function with
the wave function of the pseudo-bound state, which was calculated in the
bond state approximation with $N_{\max}=70$. Both states have approximately
the same energy; the energy of the resonance state is 0.536 MeV, while the
pseudo-bound state has the energy 0.529 MeV.%

It is worthwhile noticing that approximately such a structure of the Hoyle
state wave function has been obtained within the Complex Scaling method in
Ref. \cite{2011PhRvC..83b4301Y} and within the Fermion Molecular Dynamics in
Ref. \cite{2009FBS....45..145N}.

We determined the shape of the triangle comprised of three alpha-particles in
bound and resonance states. The average distances between clusters are
displayed in Table \ref{Tab:Shape0P12C}.%

\begin{table}[tbp] \centering
\begin{ruledtabular}
\caption{The energy, width and average distances $R_1$, $R_2$ between clusters   
for the ground states and
for the $0^{+}$ and $1^{-}$ resonance states in $^{12}$C.}%
\begin{tabular}
[c]{ccccc}
$J^{\pi}$ & $E$, MeV & $\Gamma$, keV & $R_{1}$, fm & $R_{2}$, fm\\\hline
$0^{+}$ & -11.37 & - & 3.12 & 3.60\\
& 0.68 & 2.9 & 6.95 & 8.02\\
& 5.16 & 534 & 6.43 & 7.43\\
$1^{-}$ & 3.52 & 0.21 & 6.07 & 7.00\\
\end{tabular}
\label{Tab:Shape0P12C}%
\end{ruledtabular}
\end{table}%

The shape and size of triangles for the ground and the first $0^{+}$ resonance
states are consistent with the corresponding density distributions displayed
in Figs. 8 and 9 of Ref. \cite{2012PhRvC..85c4318V}. It is interesting to
note that the shape of resonance states, shown in Table \ref{Tab:Shape0P12C},
is almost independent on the energy and total width of the resonance state,
and the structure of resonance wave functions shown in Figs.
\ref{Fig:ShellWeights0P12C} and \ref{Fig:ShellWeight12CRS}. The main
conclusion one may deduce from Table \ref{Tab:Shape0P12C} is that the average
distances between alpha-particles are rather large. The ground state of
$^{12}$C is shows a compact three-cluster configuration, as it is expected.

Having reanalyzed properties of the Hoyle state and other resonance states in
$^{12}$C, we suggest the following criteria for the Hoyle-analog states:

\begin{itemize}
\item the Hoyle-analog state is a very narrow resonance state in the
three-cluster continuum;

\item wave function of the Hoyle-analog state has large values of amplitudes
$W_{sh}$ in the internal region.
\end{itemize}

As we pointed out above, we consider the first criterion is the most important
one. We believe that the more long-lived resonance state has more chances that
 the system transits from a resonance state into a bound states, and vise
versa. It is well-known that a resonance state could substantially increase a
cross section of a processes if the total width of this resonance state is
very small. To quantify the "narrowness" of a resonance state we will
calculate the ratio $\Gamma/E$. For the original Hoyle state this ratio is
2.24 $\times$ 10$^{-7}$. Such a type of resonance states are also called as
the quasistationary states. As additional and important criterion we will use
a behavior of the weights $W_{sh}$ of oscillator shells in the wave function
of the resonance state.

Considering candidates of the Hoyle-analog states, we are also going to
check other criteria  formulated in Introduction.

\subsection{$^{9}$Be and $^{9}$B}

As was pointed out above, spectra of resonance states in $^{9}$Be and $^{9}$B
have been investigated within the present model in Refs.
\cite{2014PAN..77.555N} and \cite{PhysRevC.96.034322}. In Ref.
\cite{PhysRevC.96.034322} we have discovered several resonance states which
can be considered as the Hoyle-analog states. For completeness of the explanation
we shortly present the main results relevant to the subject of the present paper.

Energies and widths of the resonance states in $^{9}$Be and $^{9}$B presented
in Ref. \cite{PhysRevC.96.034322}  were obtained with the modified version of
the Hasegawa-Nagata potential, which is often used in numerous calculations of
two- and three-cluster structures of light nuclei. It was shown that our
three-cluster model with such the potential reproduces fairly good spectra of
resonance states in both nuclei. It was also demonstrated that the
Hasegawa-Nagata potential provides a more adequate description of resonance
states in $^{9}$Be and $^{9}$B, than the Minnesota potential (see detail in
Ref. \cite{2014PAN..77.555N}).

In Table \ref{Tab:ResonsMHN&Exp} we collect energies and widths of resonance
states in $^{9}$Be and $^{9}$B.%

\begin{table}[htbp] \centering
\begin{ruledtabular}
\caption{Spectra of resonance states of $^9$Be and $^9$B calculated within
AMHHB model with MHNP.}%
\begin{tabular}
[c]{cccccc}
\multicolumn{3}{c}{$^{9}$Be} & \multicolumn{3}{c}{$^{9}$B}\\\hline
$J^{\pi}$ & $E$, MeV & $\Gamma$, MeV & $J^{\pi}$ & $E$, MeV & $\Gamma$,
MeV\\\hline
$3/2^{-}$ & -1.574 & - & $3/2^{-}$ & 0.379 & 1.08$\times$10$^{-6}$\\
$1/2^{+}$ & 0.338 & 0.17 & $1/2^{+}$ & 0.636 & 0.48\\
$5/2^{-}$ & 0.897 & 2.36$\times$10$^{-5}$ & $5/2^{-}$ & 2.805 & 0.02\\
$5/2^{+}$ & 2.086 & 0.11 & $3/2^{+}$ & 2.338 & 2.80\\
$3/2_{2}^{-}$ & 2.704 & 2.53 & $1/2^{-}$ & 3.398 & 3.43\\
$1/2^{-}$ & 2.866 & 1.60 & $5/2^{+}$ & 3.670 & 0.42\\
$3/2^{+}$ & 4.062 & 1.22 & $3/2_{2}^{-}$ & 3.420 & 3.36\\
$7/2^{-}$ & 4.766 & 4.04 & $5/2_{2}^{-}$ & 5.697 & 5.15\\
$9/2^{+}$ & 4.913 & 1.27 & $9/2^{+}$ & 6.503 & 2.01\\
$5/2_{2}^{-}$ & 5.365 & 4.38 & $7/2^{-}$ & 6.779 & 0.90\\
\end{tabular}
\label{Tab:ResonsMHN&Exp}%
\end{ruledtabular}
\end{table}%

There is only one very narrow resonance state in each nucleus. This is the
5/2$^{-}$ resonance state in $^{9}$Be and the 3/2$^{-}$ resonance state in
$^{9}$B which is the "ground state" of the nucleus. We considered these
resonance states as candidates to the Hoyle-analog states. We also added the
1/2$^{+}$ resonance state to that list of resonance states, as they lie close
to the three-cluster threshold. Other resonance states in $^{9}$Be and $^{9}$B
have a large total width and they were disregarded.

\begin{figure}[ptbh]
\begin{center}
\includegraphics[width=\columnwidth]%
{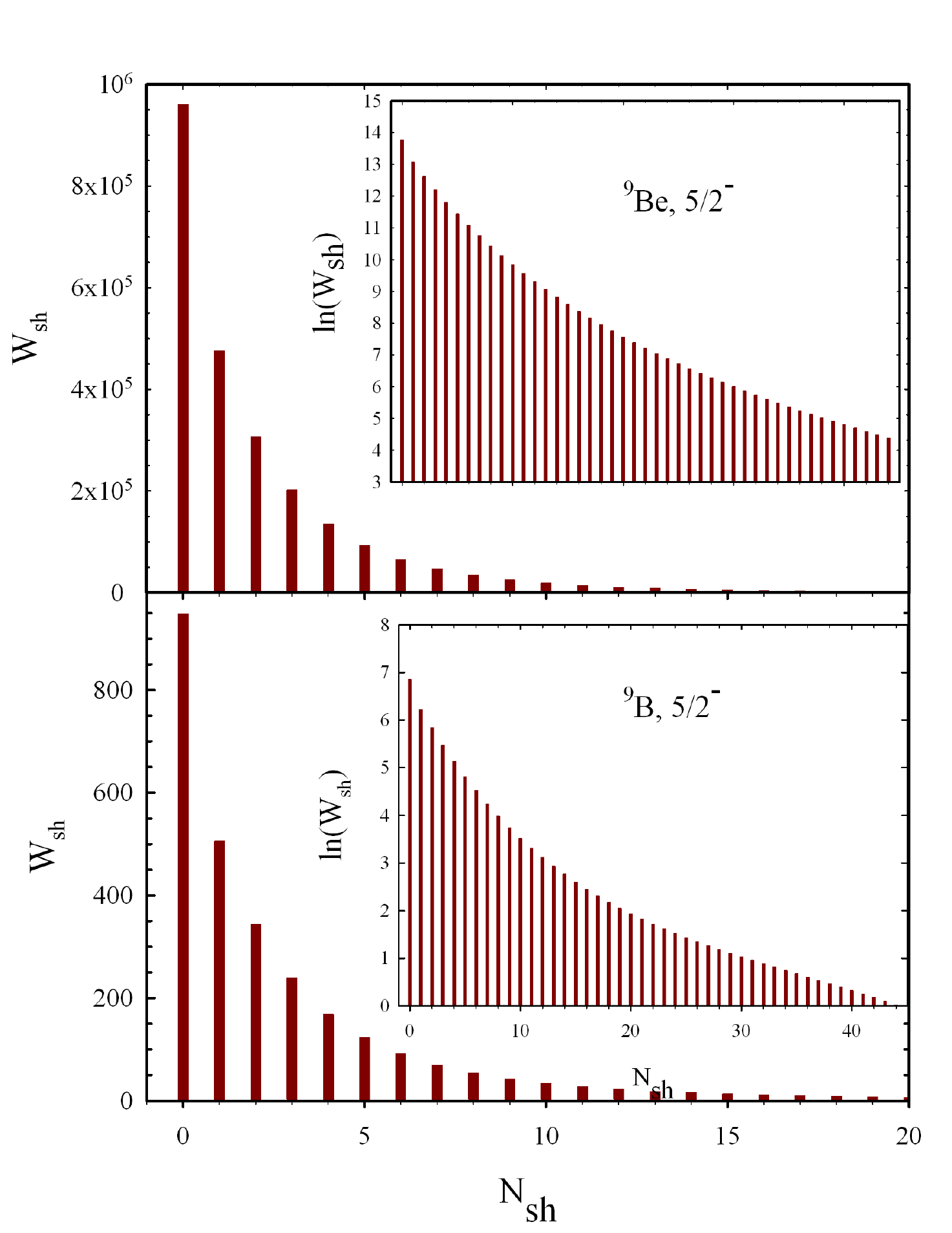}
\caption{Weight $W_{sh}$ as a function of oscillator shell $N_{sh}$ \ for the
5/2$^{-}$ resonance state in $^{9}$Be and $^{9}$B.}%
\label{Fig:ShrellWeightR52M9Be9B}%
\end{center}
\end{figure}

In Fig. \ref{Fig:ShrellWeightR52M9Be9B} we display the structure of wave
functions of the 5/2$^{-}$ resonance states in $^{9}$Be and $^{9}$B. The total
width of the 5/2$^{-}$ resonance states in $^{9}$Be is 24 eV and amplitudes of
the dominant shell weights $W_{sh}$ are of 10$^{5}$ order of magnitude. The
same resonance state in $^{9}$B is wider ($\Gamma$=18 keV) and thus amplitudes
of the dominant shell weights $W_{sh}$ are less than 1000. As one can see that
the oscillator shells with $0\leq N_{sh}<20$ give the main contribution to
the wave functions of the 5/2$^{-}$ resonance states. In Fig.
\ref{Fig:ShrellWeightR52M9Be9B}  we also display $W_{sh}$ in a logarithmic
scale to demonstrate their behavior in the internal region. Within the
internal region, wave functions are decreasing exponentially like wave
functions of bound states. Such a behavior of wave functions of the 5/2$^{-}$
resonance states in $^{9}$Be and $^{9}$B allows us to consider these resonance
states as the Hoyle-analog states.

\begin{figure}[ptbh]
\begin{center}
\includegraphics[width=\columnwidth]%
{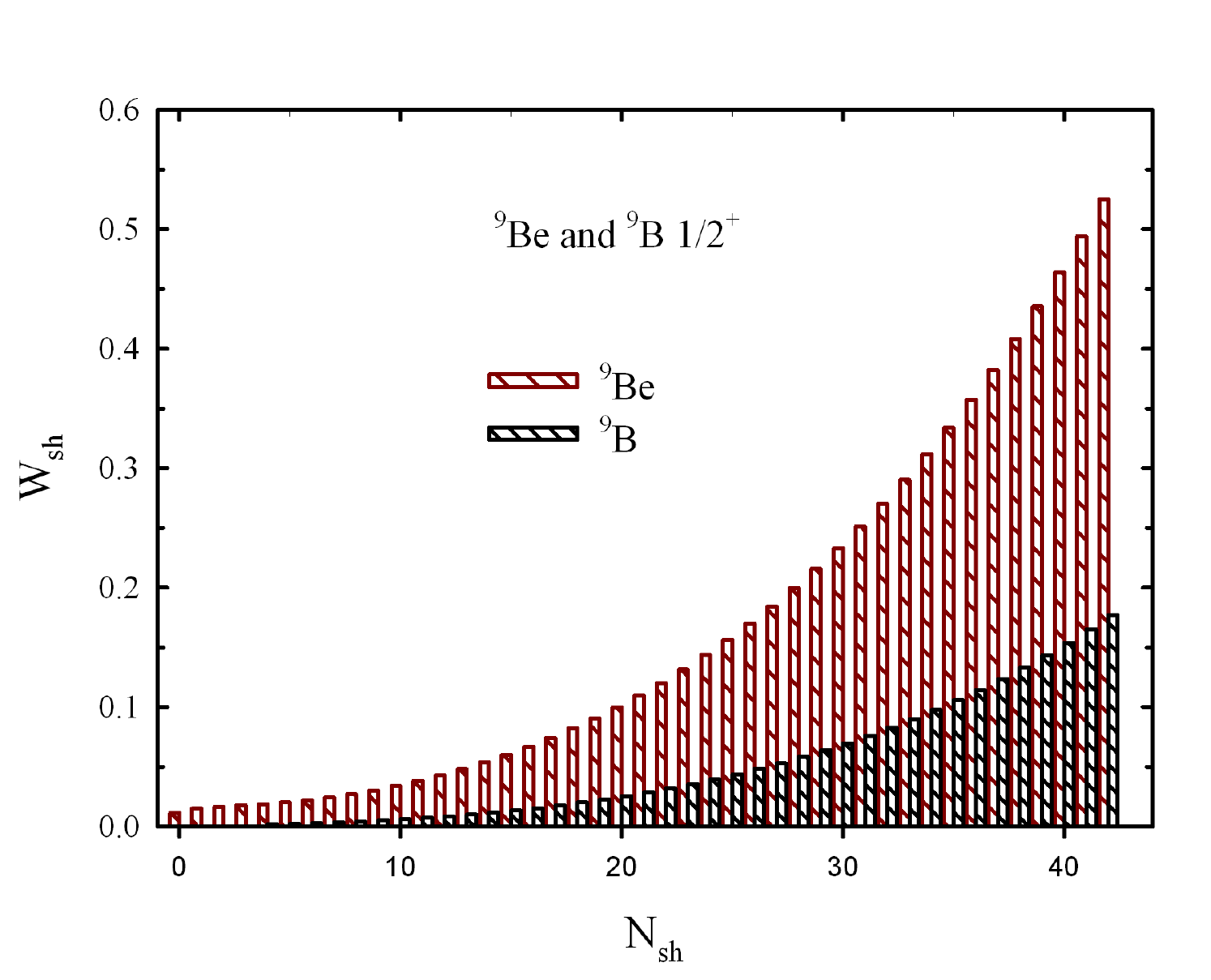}
\caption{Weights of different oscillator shells in wave functions of the
1/2$^{+}$ resonance states in $^{9}Be$ (red color) and $^{9}B$ (blue color).}%
\label{Fig:ShrellWeightR12P9Be9B}%
\end{center}
\end{figure}

In Ref. \cite{PhysRevC.96.034322} we have also considered the 1/2$^{+}$
resonance states in $^{9}$Be and $^{9}$B as possible candidates to the
Hoyle-analog states. These resonances lie very close to the three-cluster
threshold, however, the 1/2$^{+}$ resonance states are rather wide resonances
and their wave functions both in coordinate and oscillator representations
indicate a very dispersed a three-cluster configuration (see wave functions in
coordinate space in Fig. \ref{Fig:ShrellWeightR12P9Be9B}). The later is also
confirmed by the average distances $R_{1}$ and $R_{2}$.%

We also analyze all resonance states in $^{9}$Be in order to find the
Hoyle-analog state. We suggested that the 5/2$^{-}$ resonance state can be
considered as the Hoyle-analogue state as this is a very narrow resonance state.
It lives long enough and may transform to the 3/2$^{-}$ ground state of
$^{9}$Be by emitting the quadrupole gamma quanta. This reaction, which
involves the triple collision of two alpha particles and neutron and a
subsequent radiation of gamma quanta, can be considered as an additional way
for the synthesis of the $^{9}$Be nuclei.%

It was shown in Ref. \cite{PhysRevC.96.034322} that such a behavior of the
wave function (Fig. \ref{Fig:ShrellWeightR12P9Be9B}) is typical for a low-ling
resonance state with a relatively large value of the total width.

\subsection{$^{11}$B and $^{11}$C.}

Now we consider the spectra of resonance states in $^{11}$B and $^{11}$C. In
Table \ref{Tab:ResonStates11BN} we display the energy and width of resonance
states in the three-cluster $\alpha+\alpha+t$ continuum of $^{11}$B, which were
calculated in Ref. \cite{2013UkrJPh.58.544V}.%

\begin{table}[htbp] \centering
\begin{ruledtabular}
\caption{The spectrum of positive- and negative-parity resonance states in
$^{11}$B. Energy is in MeV and measured from the three-cluster 
$\alpha+\alpha+t$ threshold.}%
\begin{tabular}
[c]{cccccc}
$J^{\pi}$ & $E$, MeV & $\Gamma$, keV & $J^{\pi}$ & $E$, MeV & $\Gamma$,
keV\\\hline
$3/2^{-}$ & 0.755 & 0.58 & $1/2^{+}$ & 0.437 & 15.26\\
& 1.402 & 185.18 &  & 0.702 & 12.30\\
& 1.756 & 143.72 &  & 1.597 & 15.95\\
$1/2^{-}$ & 1.436 & 374.64 & $3/2^{+}$ & 1.147 & 1.498\\
& 1.895 & 100.95 &  & 1.367 & 8.58\\
& 2.404 & 450.07 &  & 1.715 & 41.24\\
$5/2^{-}$ & 0.583 & 5.14$\times$10$^{-4}$ & $5/2^{+}$ & 1.047 &
1.54\\
& 1.990 & 32.63 &  & 1.951 & 40.20\\
& 2.251 & 138.87 &  & 2.265 & 54.73\\
& 2.905 & 120.46 &  & 2.748 & 167.61\\
$7/2^{-}$ & 1.591 & 4.14 & $7/2^{+}$ & 1.076 & 2.04$\cdot$10$^{-2}$\\
& 1.778 & 3.04 &  & 2.119 & 26.32\\
& 2.471 & 20.18 &  & 2.536 & 100.47\\
\end{tabular}
\label{Tab:ResonStates11BN}%
\end{ruledtabular}
\end{table}%

In a small range of energies $0<E<3$ MeV we observed 26 resonance states. The
large part of these resonances are narrow resonance states with the total
width less than 50 keV. The similar picture is observed in $^{11}$C. The
energy and width of positive- and negative-parity states are shown in  Table
\ref{Tab:ResonStates11CN}. Details of these calculations can be found in Ref.
\cite{2013UkrJPh.58.544V}.%

\begin{table}[htbp] \centering
\begin{ruledtabular}
\caption{The spectrum of positive- and negative-parity resonance states in
$^{11}C$. Energy is in MeV and measured from three-cluster 
$\alpha+\alpha+ ^3$He  threshold.}%
\begin{tabular}
[c]{cccccc}
$J^{\pi}$ & $E$, MeV & $\Gamma$, keV & $J^{\pi}$ & $E$, MeV & $\Gamma$,
keV\\\hline
$3/2^{-}$ & 0.805 & 9.93$\times$10$^{-3}$ & $1/2^{+}$ & 0.906 &
162.94\\
& 1.920 & 105.08 &  & 1.930 & 59.88\\
& 2.324 & 619.76 &  & 2.679 & 86.69\\
$1/2^{-}$ & 1.142 & 0.708 & $3/2^{+}$ & 2.268 & 34.25\\
& 2.266 & 790.98 &  & 2.478 & 159.28\\
& 3.014 & 366.15 &  & 2.850 & 115.19\\
$5/2^{-}$ & \multicolumn{1}{c}{0.783} & \multicolumn{1}{c}{9.64$\times
$10$^{-5}$} & $5/2^{+}$ & 1.460 & 0.90\\
& 1.897 & 5.77 &  & 2.346 & 82.72\\
& 3.026 & 182.69 &  & 3.179 & 122.75\\
& 3.491 & 392.96 & $7/2^{+}$ & 1.765 & 7.40$\cdot$10$^{-2}$\\
$7/2^{-}$ & 2.700 & 66.63 &  & 2.542 & 8.19\\
& 3.538 & 21.18 &  & 3.237 & 119.13\\
\end{tabular}
\label{Tab:ResonStates11CN}%
\end{ruledtabular}
\end{table}%

By using the criteria for selecting the candidate to the Hoyle-analog states,
formulated above, we selected four resonance states in $^{11}$B and four
resonance states in $^{11}$C. In Table \ref{Tab:ShapeSiseRS11B11C} we display
the properties of the selected resonance states in $^{11}$B and $^{11}$C, and
compare them with some bound states. We did not include the 1/2$^{+}$
resonance state in $^{11}$C as it has a relatively large total width.%

\begin{table}[tbp] \centering
\begin{ruledtabular}
\caption{Parameters of resonance states in $^{11}$B and $^{11}$C selected as candidates
to the Hoyle-analog states. The average distances $R_1$ and $R_2$ are presented both for
resonance states and for bound states.}%
\begin{tabular}
[c]{ccccccc}
Nucleus & $J^{\pi}$ & $E$, MeV & $\Gamma$, keV & $\Gamma/E$ & $R_{1}$, fm &
$R_{2}$, fm\\\hline
& $3/2^{-}$ & -11.055 &  &  & 2.60 & 2.88\\
& $3/2^{-}$ & -5.667 &  &  & 2.90 & 3.38\\
& $3/2^{-}$ & -0.589 &  &  & 4.83 & 6.79\\
$^{11}$B & $1/2^{+}$ & 0.437 & 15.26 & 3.49$\times$10$^{-2}$ & 10.48 & 6.77\\
& $5/2^{-}$ & 0.583 & 5.14$\times$10$^{-4}$ & 8.81$\times$10$^{-7}$ & 4.71 &
7.20\\
& $3/2^{-}$ & 0.755 & 0.58 & 7.7$\times$10$^{-4}$ & 5.36 & 7.75\\
& $5/2^{+}$ & 1.047 & 1.54 & 1.47$\times$10$^{-3}$ & 4.98 & 7.47\\\hline
& $3/2^{-}$ & -9.073 &  &  & 2.64 & 2.90\\
& $3/2^{-}$ & -3.835 &  &  & 2.97 & 3.43\\
& $1/2^{+}$ & 0.906 & 162.94 &  & 10.75 & 7.08\\
$^{11}$C & $5/2^{-}$ & 0.783 & 9.64$\times$10$^{-5}$ & 1.23$\times$10$^{-7}$ &
3.20 & 3.87\\
& $3/2^{-}$ & 0.805 & 9.93$\times$10$^{-3}$ & 1.23$\times$10$^{-5}$ & 5.02 &
6.86\\
& $5/2^{+}$ & 1.460 & 0.90 & 6.16$\times$10$^{-4}$ & 5.00 & 6.69\\
\end{tabular}
\label{Tab:ShapeSiseRS11B11C}%
\end{ruledtabular}
\end{table}%

\begin{figure}
[ptbh]
\begin{center}
\includegraphics[width=\columnwidth]%
{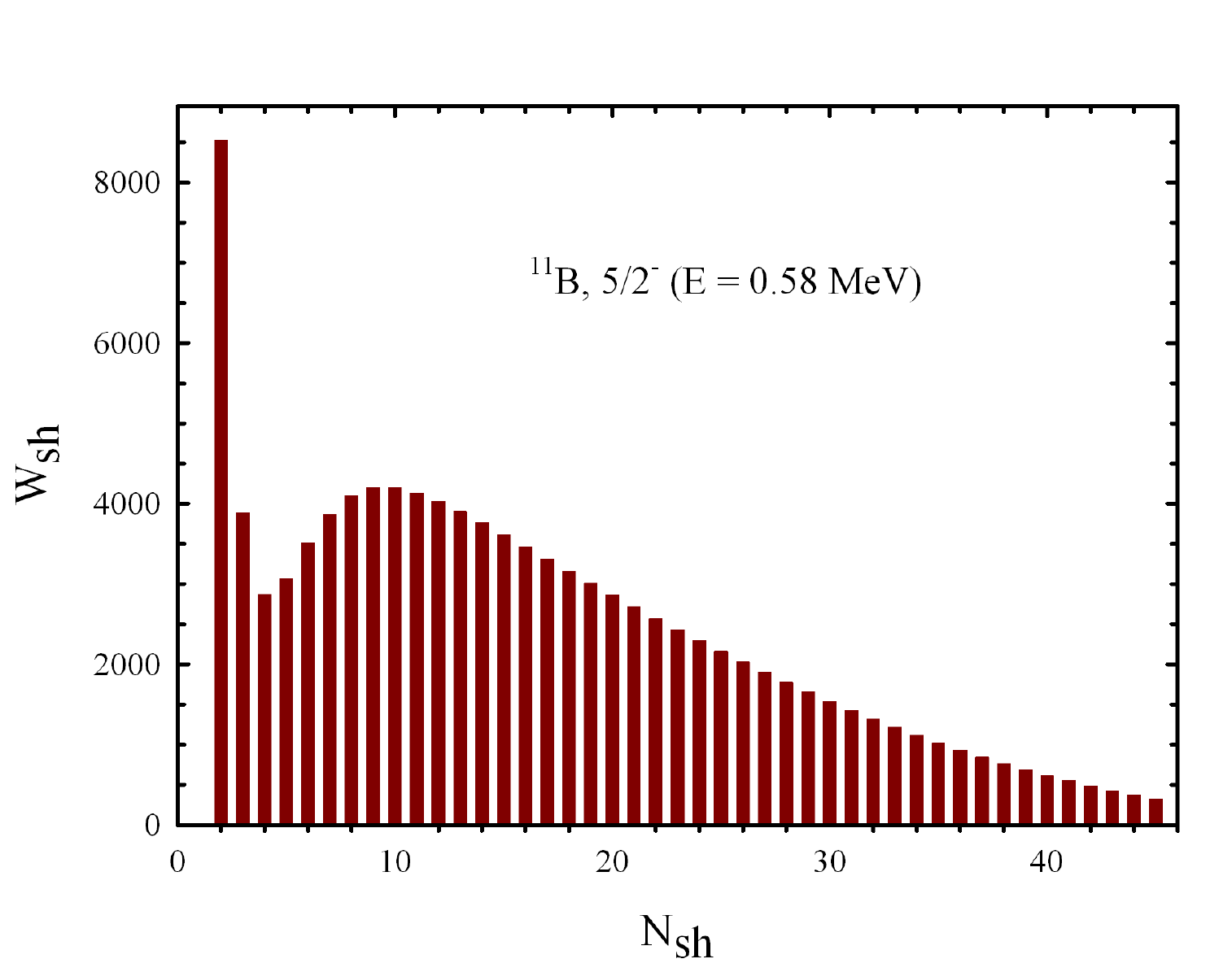}
\caption{Weights of oscillator shells in the wave functions of the 5/2$^{-}$
resonance state ($E$=0.58 MeV $\Gamma$=0.5 eV) in $^{11}$B.}%
\label{Fig:ShellWeightR52M11B}%
\end{center}
\end{figure}

\begin{figure}
[ptbh]
\begin{center}
\includegraphics[width=\columnwidth]%
{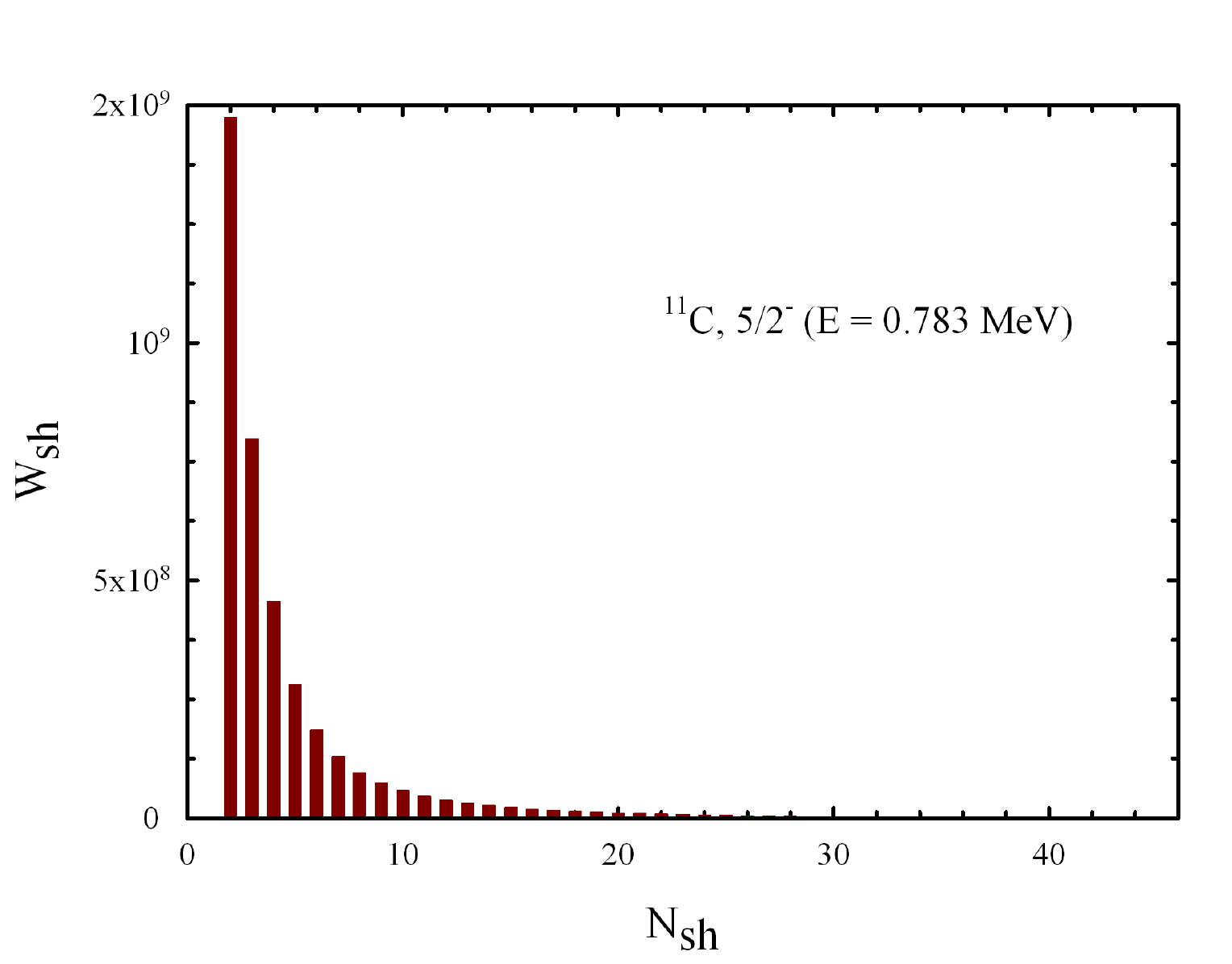}
\caption{Structure of the wave function of the very narrow 5/2$^{-}$ resonance
state in $^{11}$C.}%
\label{Fig:ShellWeightR52M11C}%
\end{center}
\end{figure}

Figs. \ref{Fig:ShellWeightR52M11B} and \ref{Fig:ShellWeightR52M11C}
demonstrating wave functions of the 5/2$^{-}$ resonance states in $^{11}$B and
$^{11}$C explicitly indicate that these resonance states can be considered as
the Hoyle analog state. Both resonance states have very large amplitudes of
weights $W_{sh}$. Structure of the wave functions of the 5/2$^{-}$ resonance
states in $^{11}$C looks like as a wave function of a bound state. These
results also show  that the average distances between clusters $R_{1}$ and
$R_{2}$ in these resonance states are very close to average distances for
bound states, for instance, for the first excited 3/2$^{-}$ state in $^{11}$C.
 From the average distances $R_{1}$ and $R_{2}$ for the resonance states in
$^{11}$B and $^{11}$C in Table \ref{Tab:ShapeSiseRS11B11C}, we see that the
most narrow 5/2$^{-}$ resonance states in $^{11}$C has the most compact
configuration of three clusters. Contrary to this resonance state, the
narrowest 5/2$^{-}$ resonance states at $E$=0.583 MeV in $^{11}$B, the total
width of which is five time larger than the width of the 5/2$^{-}$ resonance
states in $^{11}$C, has rather dispersed structure with the average distance
between alpha particles equals to 7.2 fm.%

%

\begin{figure}
[ptbh]
\begin{center}
\includegraphics[width=\columnwidth]%
{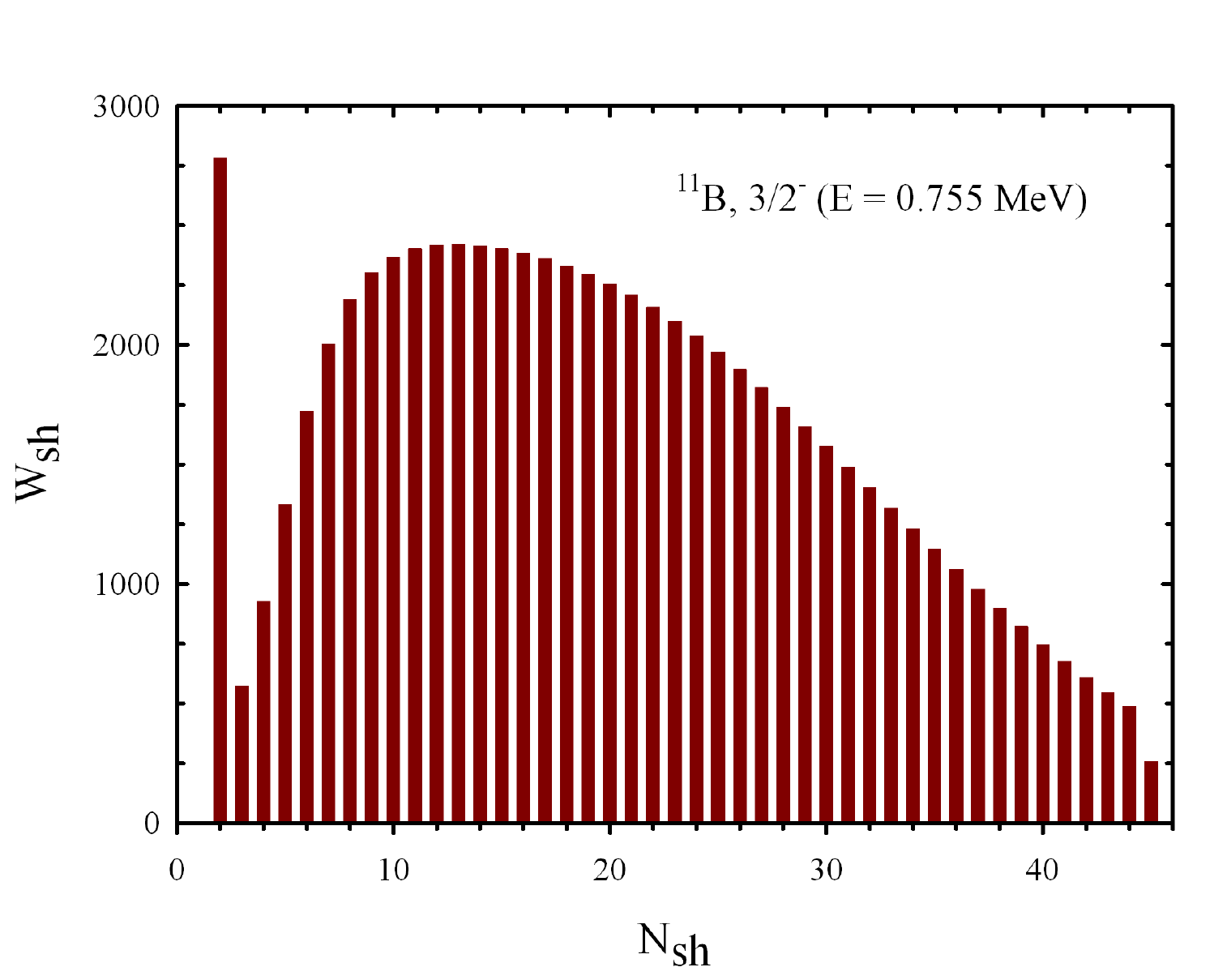}
\caption{The shape of the wave function of the 3/2$^{-}$ resonance state in
$^{11}$B.}%
\label{Fig:ShellWeightR32M11B}%
\end{center}
\end{figure}

\begin{figure}
[ptbh]
\begin{center}
\includegraphics[width=\columnwidth]%
{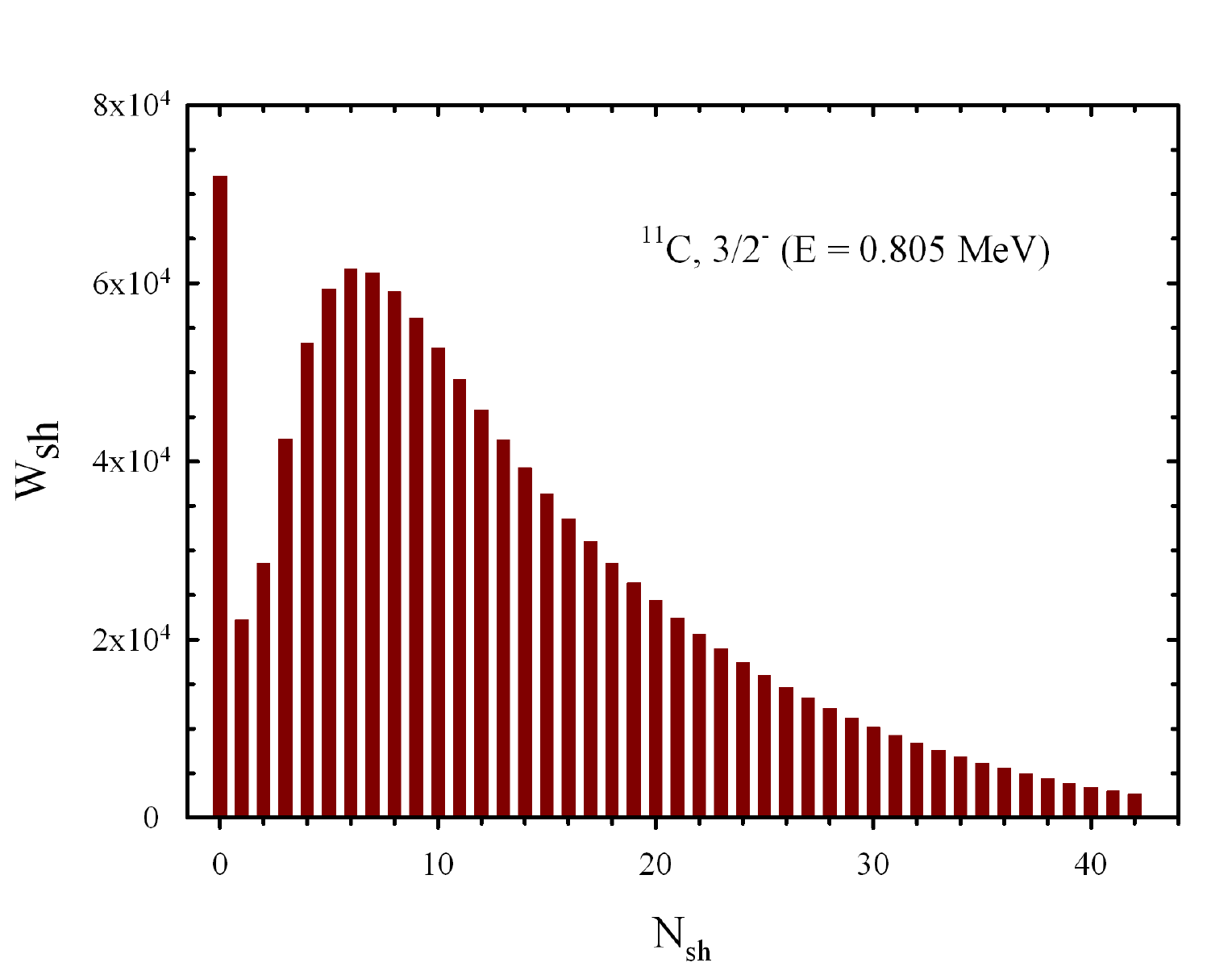}
\caption{The shape of wave function of the 3/2$^{-}$ resonance state in
$^{11}$C.}%
\label{Fig:ShellWeightR32M11C}%
\end{center}
\end{figure}

It is interesting to note that the Coulomb interaction makes the 5/2$^-$
resonance state in $^{11}$C more narrower than that in $^{11}$B. As one may
expect, it also increase the energy of the resonance in $^{11}$C comparing to
its position in $^{11}$B. The same picture is observed for the 3/2$^{-}$
resonance states in $^{11}$B and $^{11}$C. Larger the Coulomb barrier in
$^{11}$C leads to the very large amplitudes of $W_{sh}$ for the 3/2$^{-}$
resonances state. One can compare amplitudes of $W_{sh}$ for 3/2$^{-}$
resonance states in $^{11}$B and $^{11}$C in Figs.
\ref{Fig:ShellWeightR32M11B}\ and \ref{Fig:ShellWeightR32M11C}, respectively.%

%

We do not show wave functions of the 1/2$^{+}$ resonance states in $^{11}$B
and $^{11}$C here as they are very similar to the wave functions of these resonance
states in $^{9}$Be and $^{9}$B. Moreover, the shape of three-cluster triangles
in those pairs of nuclei is also similar.


As we pointed out in Introduction, there are few publications which devoted to
the Hoyle-analog states in $^{11}$B and $^{11}$C. In Refs.
\cite{2007PhRvC..75b4302K} and \cite{2011JPhCS.321a2009K}, the spectra of 
$^{11}$B and $^{11}$C have been obtained within the
antisymmetrized molecular dynamics (AMD).  The excited states have been 
treated as  bound states, it mean that the widths and energies of these 
states with respect to the three-cluster thresholds were not determined. By 
analyzing the  probability of electromagnetic transitions, authors came to 
the conclusion that the 3/2$^-$ excited states have a dilute cluster structure 
$\alpha+\alpha+t$ and $\alpha+\alpha+^3$He, and thus can be considered as the 
Hoyle-analog states. It was also claimed by the authors, that the 5/2$^-$ states 
have no a well-developed cluster structure and therefore cannot be considered 
as the Hoyle-analog states.  

In Refs. \cite{2010PhRvC..82f4315Y} and
\cite{2012PThPS.196..388Y}  resonance states in two- and three-body continuum
of $^{11}$B and $^{11}$C have been determined with the complex scaling method.
The 3/2$^{-}$ resonance state is located bellow the $\alpha+\alpha+t$
threshold has a compact cluster configuration, as was shown by the authors of
Refs. \cite{2010PhRvC..82f4315Y, 2012PThPS.196..388Y},  and
therefore was not considered as a candidate to the Hoyle-analog states. The
wave function of the 1/2$^{+}$ resonance state which has "the gas-like
structure with a large nuclear radius", as stressed by the authors, and thus
can be considered as the Hoyle-analog state. It is interesting to note that
the parameters ($E$=0.75 MeV, $\Gamma$=190 keV) of the 1/2$^{+}$ resonance
state, determined in Refs. \cite{2010PhRvC..82f4315Y, 2012PThPS.196..388Y}, 
are rather different to those ($E$=0.44 MeV,
$\Gamma$=15 keV) displayed in the present paper. This difference can be
ascribed to the different types of nucleon-nucleon potentials involved in
these two calculations. The 1/2$^{+}$ resonance state, obtained in our
calculations, has also a large nuclear radius, however is not considered as
the Hoyle-analog state in our criterion.

\subsection{$^{10}$B}

In Table \ref{Tab:Resonan10B} we show the three-cluster resonance states in
$^{10}$B calculated with the MP. Details of these calculations can be found in
Ref. \cite{2014UkrJPh..59.1065N},  where the spectrum of bound states of
$^{10}$B has been discussed. Here, we use the same input parameters to
calculate spectrum of resonance states in the three-cluster $\alpha+\alpha+d$
continuum. As we can see in Table \ref{Tab:Resonan10B}, there are a few narrow
resonance states which can be considered as the Hoyle-analog states. Three
resonance states have the total width less than 12 keV and the ratio
$\Gamma/E$ does not exceed 11.5$\times$10$^{-3}$.%

\begin{table}[tbp] \centering
\begin{ruledtabular}
\caption{Parameters of resonance states in $^{10}B$. The average distances 
$R_1$ and $R_2$ are calculated for the candidates to the Hoyle-analog states.}%
\begin{tabular}
[c]{cccccc}
$J^{\pi}$ & $E$, MeV & $\Gamma$, keV & $\Gamma/E$ & $R_{1}$, fm & $R_{2}$,
fm\\\hline
$1^{+}$ & 0.604 & \multicolumn{1}{c}{232.30} & 0.384 &  & \\
& 0.987 & \multicolumn{1}{c}{7.08} & 7.17$\times$10$^{-3}$ & 6.67 & 10.67\\
& 1.536 & \multicolumn{1}{c}{196.36} & 0.128 &  & \\
$2^{+}$ & 1.055 & \multicolumn{1}{c}{12.063} & 11.43$\times$10$^{-3}$ &
6.64 & 10.83\\
& 2.810 & \multicolumn{1}{c}{170.74} & 60.76$\times$10$^{-3}$ &  & \\
$3^{+}$ & 1.062 & \multicolumn{1}{c}{11.73} & 11.05$\times$10$^{-3}$ &
6.43 & 10.35\\
& 2.202 & \multicolumn{1}{c}{526.47} & 0.239 &  & \\
$1^{-}$ & 1.100 & \multicolumn{1}{c}{76.75} & 69.77$\times$10$^{-3}$ &
9.31 & 10.84\\
& 1.820 & \multicolumn{1}{c}{562.71} & 0.309 &  & \\
\end{tabular}
\label{Tab:Resonan10B}%
\end{ruledtabular}
\end{table}%

In Table \ref{Tab:Resonan10B} we also show the average distances between
interacting clusters. It is necessary to recall that $R_{2}$ stands for the
distance between two alpha-particles, and $R_{1}$ denotes the distance
between the deuteron and the center of mass of two alpha-particles. It is
interesting to compare the average distance between clusters for resonance
states with those for the bound states. For the ground $3^{+}$ state we have
obtained $R_{1}$ =2.60 fm  and $R_{2}$ = 3.10 fm. This is a compact
configuration despite that the binding energy is -5.95 MeV, accounted from the
three-cluster threshold $\alpha+\alpha+d$, is not very small. The first
excited $3^{+}$ state is a weakly bound state as its energy is -0.95 MeV,
however it is also a rather compact configuration with the average distances
$R_{1}$ = 4.07 and $R_{1}$ = 5.35 fm. As we see in Table
\ref{Tab:Resonan10B}, all resonance states  selected as the candidates to the
Hoyle-analog states  have a dispersed configuration with a large distance
between alpha particles.

\begin{figure}
[ptbh]
\begin{center}
\includegraphics[width=\columnwidth]%
{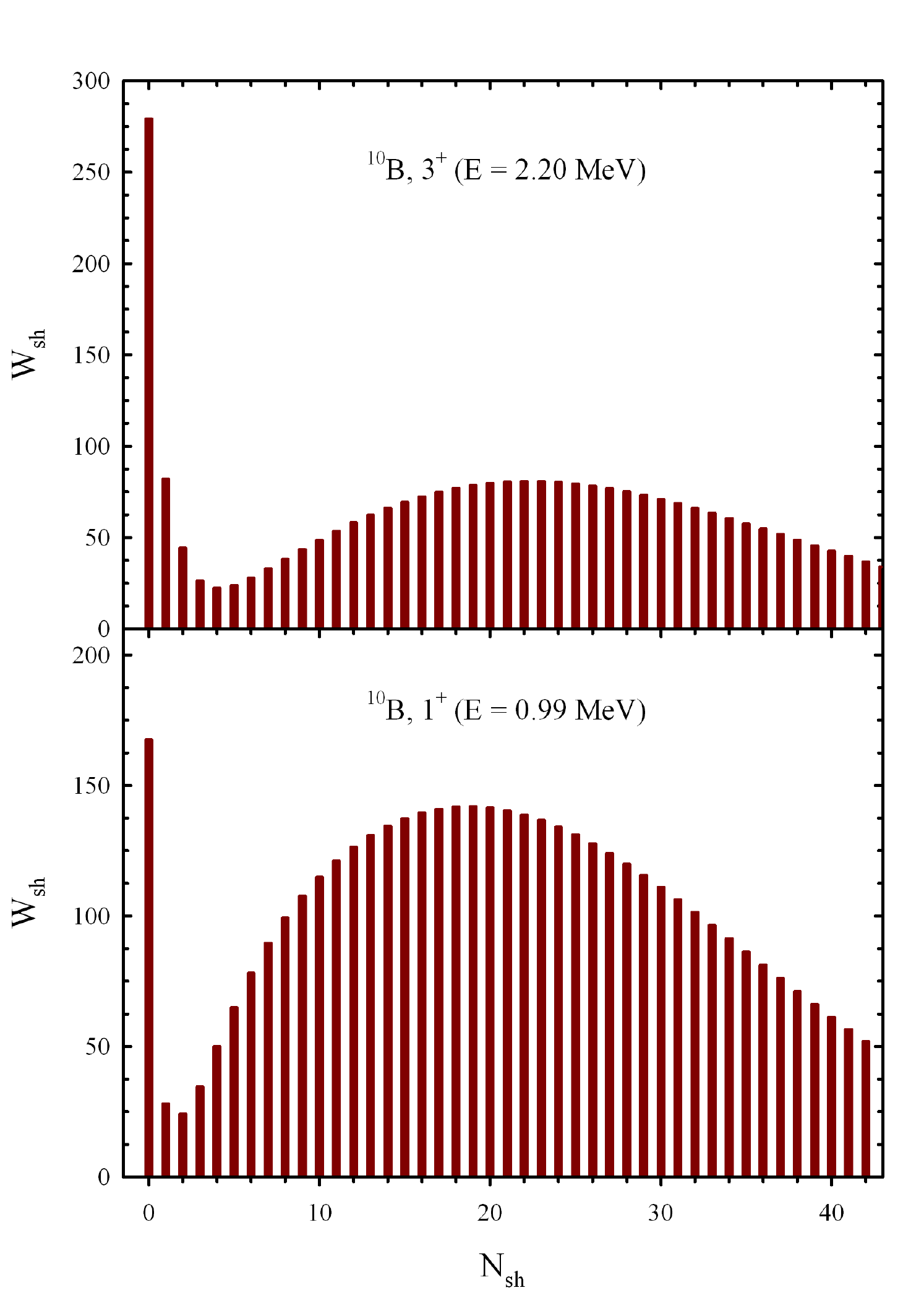}
\caption{Weights of different oscillator shells in wave functions of the
$3^{+}$ and $1^{+}$ resonance states in $^{10}$B.}%
\label{Fig:ShellWeights10B}%
\end{center}
\end{figure}

Let us turn our attention to the wave functions of the selected resonance
states. In Fig. \ref{Fig:ShellWeights10B} we display shell weights in wave
functions of the narrow $3^{+}$ and $1^{+}$ resonance states in $^{10}$B.
These resonance state have the smallest total width among all resonances in
$^{10}$B. One notices, that the compact three-cluster configuration ($N_{sh}$
=0) has a relatively large contribution to these wave functions. The shapes of
the curves are similar to the shape of the Hoyle state (Fig.
\ref{Fig:ShellWeights0P12C}), however the amplitudes are much more smaller.%

We assume that the interplay of the attractive potential, created by the
central and spin-orbital parts of the nucleon-nucleon interaction, and
repulsive potential, formed by the Coulomb interaction, does not create a
favorable situation for very narrow resonance states in $^{10}$Be.

\section{Conclusion \label{Sec:Conclusion}}

We have performed a systematic investigation of  the three-cluster resonance
states in light nuclei $^{9}$Be, $^{9}$B, $^{10}$B, $^{11}$B, $^{11}$C and
$^{12}$C. These nuclei have been considered to have a three-cluster structure
composed of two alpha particles and an $s$-shell nucleus. A microscopic three-cluster
 model was applied to search and to study resonance states
 embedded in the three-cluster continuum. This model imposes proper boundary
conditions by employing hyperspherical coordinates and hyperspherical
harmonics. Having reanalyzed properties of the Hoyle state, we formulated
criteria for the Hoyle-analog states. Among these resonances, we have found
the Hoyle-analog states in these nuclei. The Hoyle-analog states are created
by a collision of two alpha-particles and a neutron, proton, triton and
nucleus $^{3}$He. These resonance states have very small width. We discussed
an alternative way for the synthesis of light nuclei in a triple collision, in
the same manner as was suggest by F. Hoyle for $^{12}$C. We found several
resonance states having the total width of a few eV. Most of the
obtained resonance states have the width of few dozens of keV.

In Table \ref{Tab:HoyleStates} we collect the parameters of the Hoyle-analog
states in light nuclei under consideration. Results presented in this Table 
allow us to formulate the new criteria for selecting the Hoyle-analog states.  
A three-cluster resonance state can be treated as the Hoyle-analog state if  
the ratio $E/\Gamma <$ 2$\times$10$^{-3}$ for this resonance state. 

\begin{figure}
[ptbh]
\begin{center}
\includegraphics[width=\columnwidth]%
{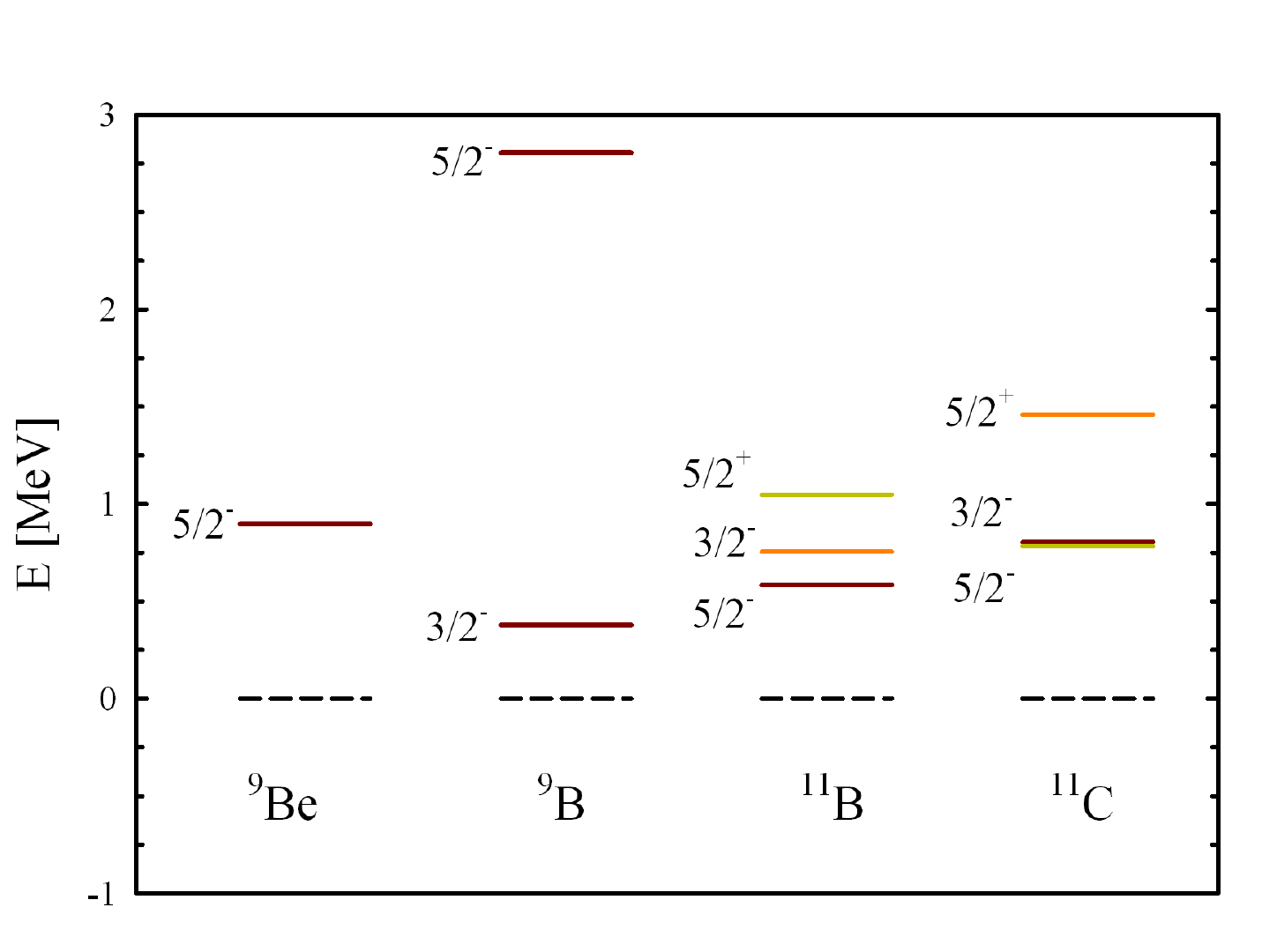}
\caption{Spectrum of the Hoyle-analog states in $^{9}$Be, $^{9}$B, $^{11}$B
and $^{11}$C.}%
\label{Fig:HoyleAnStates}%
\end{center}
\end{figure}

Figure \ref{Fig:HoyleAnStates}
visualizes the results presented in Table \ref{Tab:HoyleStates}. This
figure explicitly demonstrates effects of the Coulomb interaction on the energy of
three-cluster resonance states in mirror nuclei $^{9}$Be and $^{9}$B and 
$^{11}$B and $^{11}$C. One can see that the
Coulomb interaction has a more stronger impact on the position of the 5/2$^{-}$ 
resonance states  in $^{9}$Be and $^{9}$B than on
the position of the 3/2$^{-}$, 5/2$^{-}$ and 5/2$^{+}$ 
resonance states in $^{11}$B and $^{11}$C.%

\begin{table}[tbp] \centering
\begin{ruledtabular}
\caption{Parameters of the Hoyle-analog states in light nuclei  $^{9}$Be,
$^{9}$B, $^{11}$B and $^{11}$C.}%
\begin{tabular}
[c]{cccccc}
Nucleus & Configuration & $J^{\pi}$ & $E$, MeV & $\Gamma$, keV & $\Gamma
/E$\\\hline
$^{9}Be$ & $\alpha+\alpha+n$ & $5/2^{-}$ & 0.897 & 2.36$\cdot$10$^{-2}$ &
2.63$\cdot$10$^{-5}$\\
$^{9}B$ & $\alpha+\alpha+n$ & $3/2^{-}$ & 0.379 & 1.08$\cdot$10$^{-3}$ &
2.84$\cdot$10$^{-6}$\\
&  & $5/2^{-}$ & 2.805 & 18.0$\cdot$10$^{-3}$ & 6.42$\cdot$10$^{-6}$\\
$^{11}B$ & $\alpha+\alpha+^{3}H$ & $5/2^{-}$ & 0.583 & 5.14$\cdot$10$^{-4}$ &
8.87$\cdot$10$^{-7}$\\
&  & $3/2^{-}$ & 0.755 & 0.58 & 7.70$\times$10$^{-4}$\\
&  & $5/2^{+}$ & 1.047 & 1.54 & 1.47$\times$10$^{-3}$\\
$^{11}C$ & $\alpha+\alpha+^{3}He$ & $5/2^{-}$ & 0.783 & 9.64$\cdot$10$^{-5}$ &
1.23$\cdot$10$^{-7}$\\
&  & $3/2^{-}$ & 0.805 & 9.93$\cdot$10$^{-3}$ & 1.23$\cdot$10$^{-5}$\\
&  & $5/2^{+}$ & 1.460 & 0.90 & 6.16$\times$10$^{-4}$\\
\end{tabular}
\label{Tab:HoyleStates}%
\end{ruledtabular}
\end{table}%

In Table \ref{Tab:Resonan0P} we collect all resonance states for the total
momentum $J$ and positive parity, where the zero value of the total orbital
momentum ($L=0$) is dominant. This is the case for $^{9}$Be, $^{9}$B, $^{11}$B
and $^{11}$C. The continuous spectrum states with $L=0$ can be interpreted as
a head-on collision of the third cluster with the $^{8}$Be nucleus being in
the $0^{+}$ state. As one can see, all these resonance states lie rather close
to the three-cluster threshold and they are fairly wide, as the total widths are
 $\Gamma$=15 keV and more. Therefor they cannot be considered as the
Hoyle-analog states.%

\begin{table}[tbp] \centering
\begin{ruledtabular}
\caption{The energy and width of resonance states which are created by 
the three-cluster configuration with the total orbital momentum $L=0$.}%
\begin{tabular}
[c]{cccc}
Nucleus & $J^{\pi}$ & $E$, MeV & $\Gamma$, keV\\\hline
$^{9}$Be & $1/2^{+}$ & 0.338 & 168\\
$^{9}$B & $1/2^{+}$ & 0.636 & 477\\
$^{10}$B & $1^{+}$ & 0.604 & 232\\
$^{11}$B & $1/2^{+}$ & 0.437 & 15\\
$^{11}$C & $1/2^{+}$ & 0.906 & 163\\
\end{tabular}
\label{Tab:Resonan0P}%
\end{ruledtabular}
\end{table}%

\begin{figure}[ptbh]
\begin{center}
\includegraphics[width=\columnwidth]%
{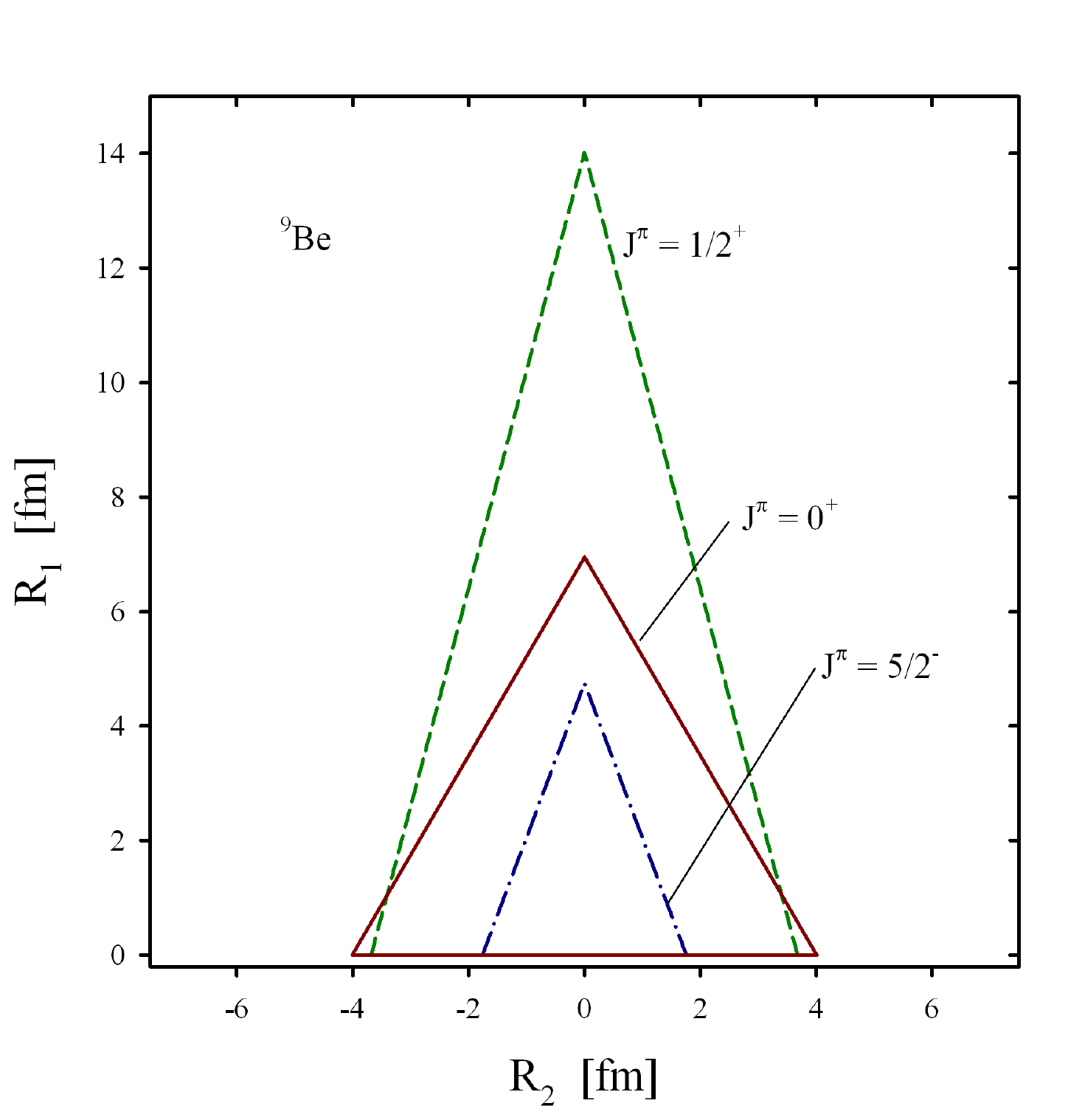}
\caption{Shape of triangles for some selected resonance states in $^{9}$Be and
the Hoyle state ($J^{\pi}=0^{+}$).}%
\label{Fig:ShapeRS9B}%
\end{center}
\end{figure}

\begin{figure}[ptbh]
\begin{center}
\includegraphics[width=\columnwidth]%
{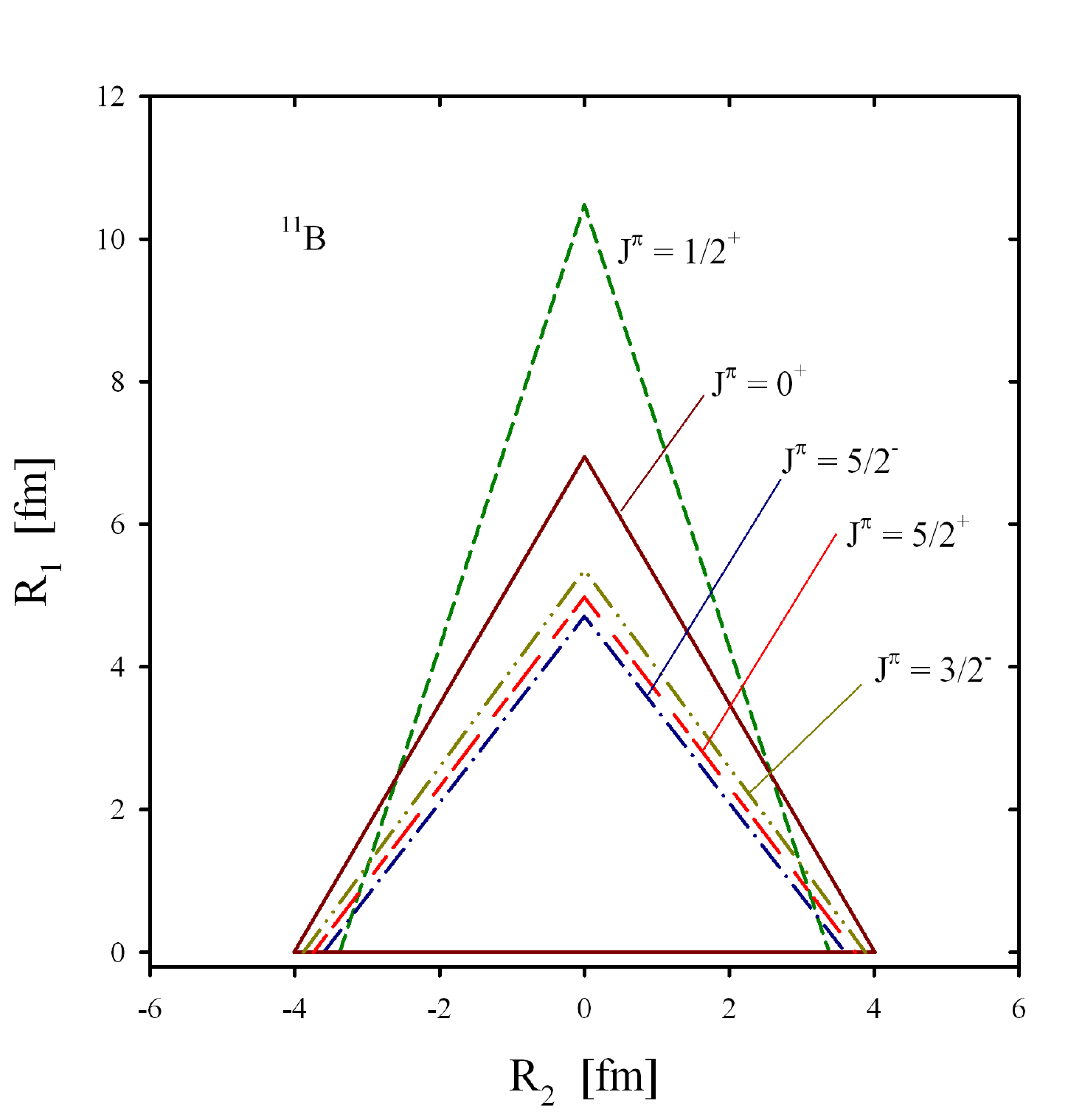}
\caption{Shape of triangles of some resonance states in $^{11}$B and the Hoyle
resonance state ($J^{\pi}=0^{+}$).}%
\label{Fig:ShapeRS11B}%
\end{center}
\end{figure}

\begin{figure}[ptbh]
\begin{center}
\includegraphics[width=\columnwidth]%
{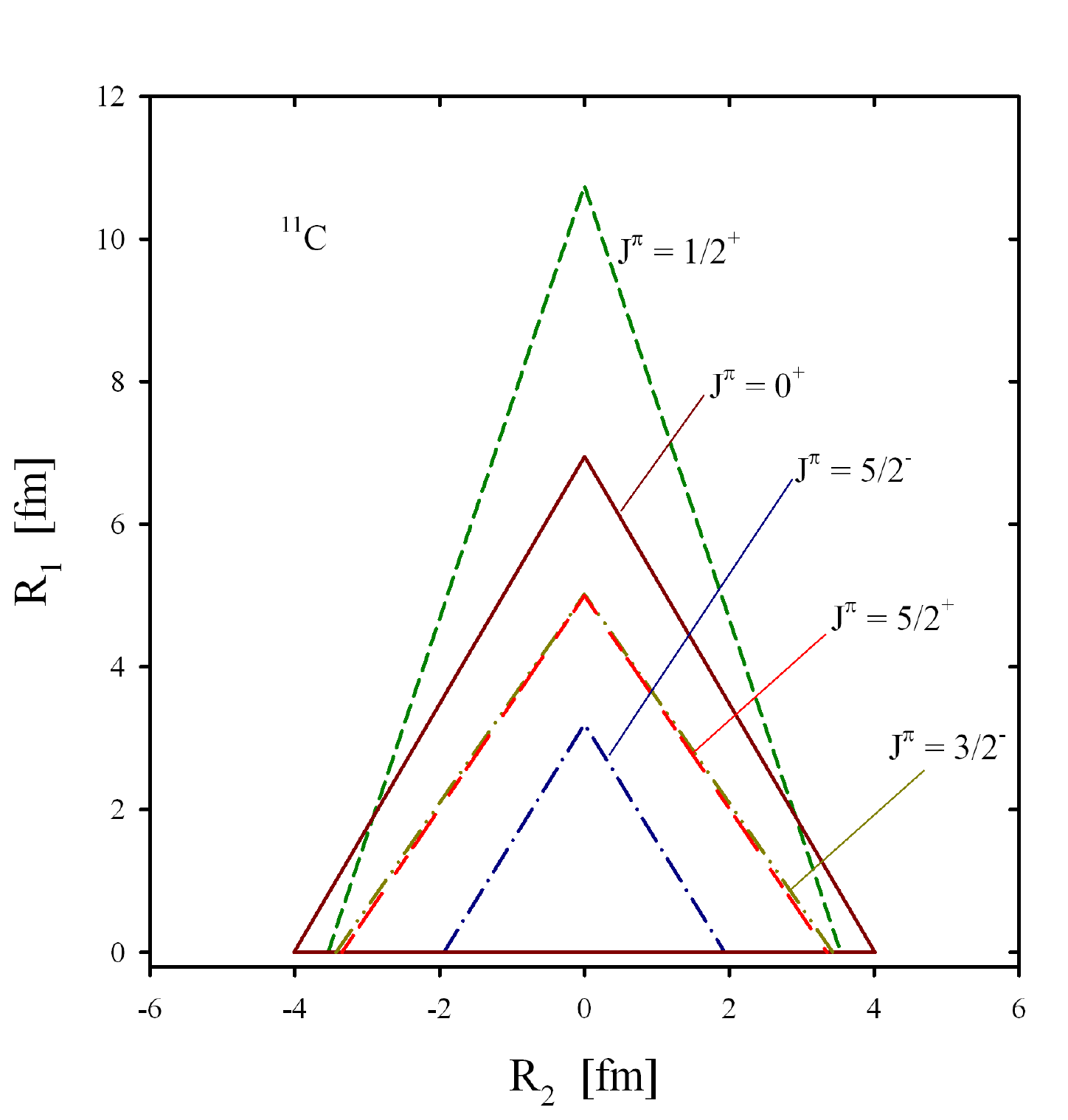}
\caption{Shape of triangles of some resonance states in $^{11}$C compared to
the Hoyle resonance state ($J^{\pi}=0^{+}$).}%
\label{Fig:ShapeRS11C}%
\end{center}
\end{figure}

Figures \ref{Fig:ShapeRS9B}, \ref{Fig:ShapeRS11B} and \ref{Fig:ShapeRS11C} of
average distances $R_{1}$ and $R_{2}$ demonstrate the most probable shapes of
triangles of three-cluster resonance states in $^{9}$Be, $^{11}$B and  $^{11}%
$C, respectively. In all these Figures we also show the triangle comprised by
three alpha-particles in the Hoyle resonance state in $^{12}$C. The 1/2$^{+}$
resonance states  in $^{9}$B, $^{11}$B and $^{11}$C has very large
triangles where a neutron, triton and $^{3}$He nucleus are far away from two
alpha-particles. The Hoyle-analog states in these nuclei have a triangle
comparable with the shape of the Hoyle-state and in some cases (for example,
for $J^{\pi}$ = 5/2$^{-}$) they are more compact.

%
%
%
%
%
%
\begin{figure}[ptbh]
\begin{center}
\includegraphics[width=\columnwidth]%
{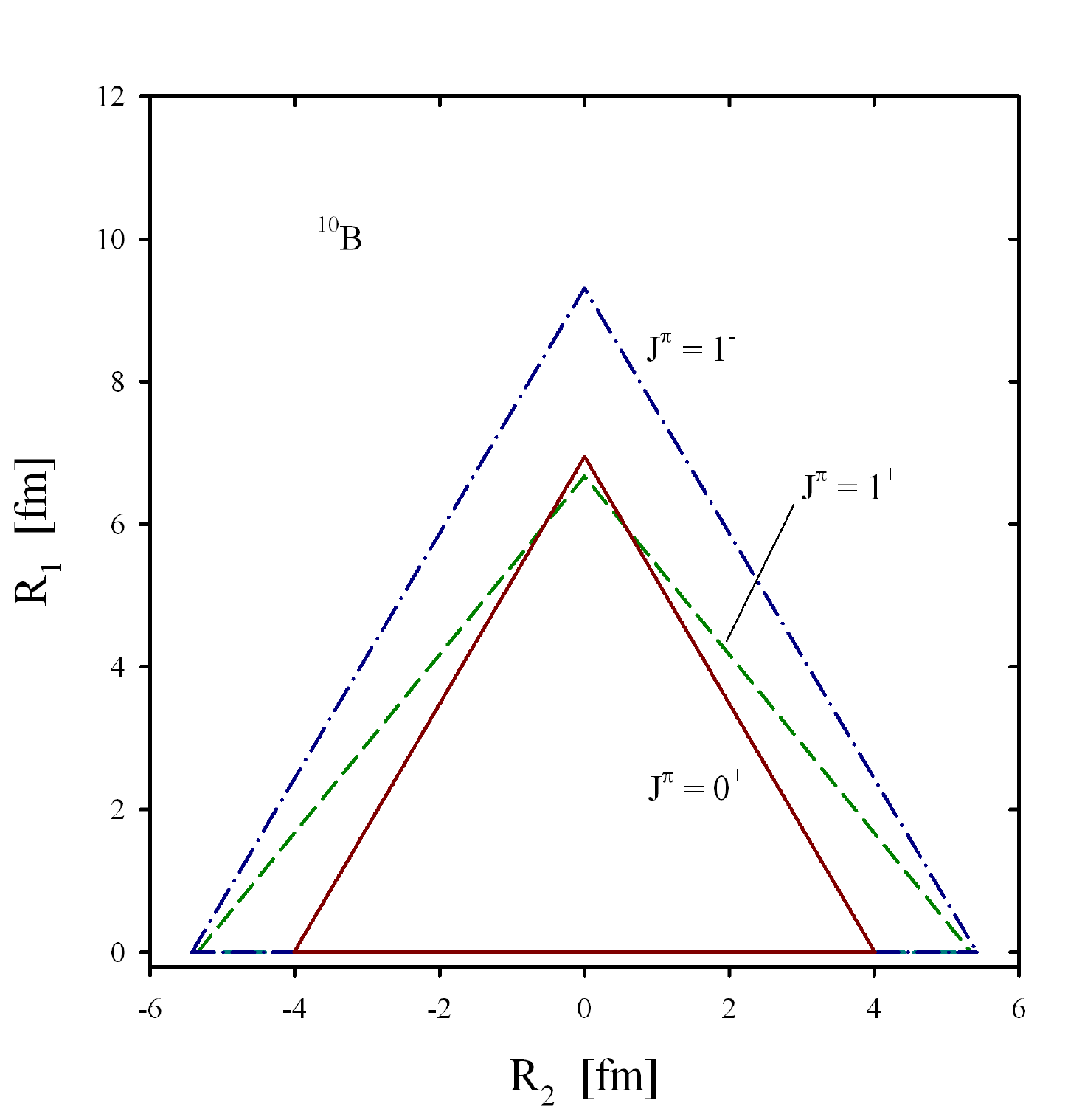}
\caption{The most probable shape of three-cluster triangles for most narrow
resonance states in $^{10}$B and compared with the Hoyle state ($J^{\pi}%
=0^{+}$).}%
\label{Fig:ShapeRS10B}%
\end{center}
\end{figure}
%
Figure \ref{Fig:ShapeRS10B} demonstrates the shape of triangles for resonance
states in $^{10}$B. They are the narrowest resonance states, however the
amplitudes of  $W_{sh}$, as was shown above, are fairly small and the ratios
$E/\Gamma$ for these states are large. They don't match our criteria for the
Hoyle-analog states. As we see, the distance between alpha-particles is
greater than this distance in the $0^{+}$ resonance state in $^{12}$C and all
other nuclei considered in the present paper. To this end, the 1/2$^{+}$
resonance states in $^{9}$Be, $^{9}$B, $^{11}$B and $^{11}$C which are
considered as candidates to the Hoyle states and did not match our criteria,
have the average distance between two alpha-particles comparable with the
Hoyle state, meanwhile the average distance of the third cluster (neutron,
proton, triton and $^{3}$He, respectively) to the center of mass of two alpha
particles is very large.

\section*{Acknowledgement}

This work was supported in part by the Program of Fundamental Research of the
Physics and Astronomy Department of the National Academy of Sciences of
Ukraine (Project No. 0117U000239) and by the Ministry of Education and Science
\ of the Republic of Kazakhstan, Research Grant  IRN: AP 05132476.

\end{document}